\documentclass[twocolumn,showpacs,assymb,amsmath]{revtex4}
\usepackage{mathrsfs}
\usepackage{txfonts}
\usepackage{graphicx}
\usepackage{amssymb}
\usepackage{amsmath}
\usepackage{color}

\begin{document}

\title{A multi-pathway  model for Photosynthetic reaction center}

\author{M. Qin$^{1,2}$, H. Z Shen$^{1,2}$, and X. X. Yi$^2$
\footnote{Corresponding address: yixx@nenu.edu.cn}}
\affiliation{$^1$School of Physics and Optoelectronic Technology\\
Dalian University of Technology, Dalian 116024, China\\
$^2$Center for Quantum Sciences and School of Physics, Northeast
Normal University, Changchun 130024, China}

\begin{abstract}
Charge separation in light-harvesting complexes occurs in a pair of
tightly coupled chlorophylls   at the heart of photosynthetic
reaction centers of both plants and bacteria. Recently it has been
shown that quantum coherence can, in principle, enhance the
efficiency of a solar cell, working like a quantum heat engine
(QHE). Here, we propose a biological quantum heat engine (BQHE)
motivated by Photosystem {\rm II} reaction center (PS{\rm II} RC) to
describe the charge separation. Our model mainly considers two
charge-separation pathways more than  that  in the published
literature. The two pathways can interfere via cross-couplings and
work together to enhance the charge-separation yields. We explore
how these cross-couplings increase the current and voltage of the
charge separation and discuss  the advantages of multiple pathways
 in terms  of current and power.  The robustness of
the BQHE against the charge recombination in natural PS{\rm II} RC
and dephasing induced by environments is also  explored, and
extension from two pathways to multiple pathways is made. These
results suggest that nature-mimicking architectures with engineered
multiple pathways for charge separations might be better for
artificial solar energy devices.
\end{abstract}

 \maketitle
%42.50.Nn: Quantum optical phenomena in absorbing, amplifying,
%          dispersive and  conducting media; cooperative phenomena in quantum optical systems
%84.60.Jt: Photoelectric conversion
%42.50.Gy: Effects of atomic coherence on propagation, absorption, and
%          amplification of light; electromagnetically induced transparency and absorption
%82.39.Jn: Charge (electron, proton) transfer in biological systems
\section{Introduction}
Photosynthesis begins with the absorption of a photon which creates
an excited state on a pigment molecule. The excitation is
transferred between the pigments of light-harvesting complexes until
it arrives at a reaction center (RC) in the pigment-protein complex,
where the photon energy is used for the later dark stages of charge
separation and conversion of energy from physical into  chemical
one. The efficiency of the energy transfer   is very high, with a
near unity yield \cite{Blankenship2002}. This sparks  the
long-standing and increasing interest in the understanding of the
physics behind the energy conversion within photosynthesis
\cite{Amerongen2000}. Recently, much attention has been paid to the
role of quantum coherence: growing experimental evidence
\cite{Brixner2005434,Engel2007446,Calhoun2009113,
Panitchayangkoon2010107,Hayes201012,Collini2010463,Harel2012109} and
theoretical models
\cite{Plenio200810,Mohseni2008129,Olaya200878,Ishizaki2009106,Caruso2009131,
Rebentrost200911,Scholes20101,Chin201012,Lloyd201012,Strumpfer20123,Dong20121,Chin20139,Rey20134}
reveal that quantum coherence contributes beneficially to the high
efficiency. Understanding the underlying mechanism of such natural
system can assist us in designing novel nanofabricated structures
for quantum transport and optimized solar cells.

Viewing  the photosynthetic reaction center as a quantum heat engine
(QHE), the authors of Ref.\cite{Dorfman2013110} analyzed  the charge
separation in light-harvesting complexes. This treatment bridges the
two seemingly unrelated effects attributed to quantum coherence in
natural(photosynthesis) and artificial(photovoltaics)
light-harvesting systems.  The common ground between photovoltaics
and photosynthesis has also been investigated recently in
\cite{Fingerhut201012,Blankenship2011332}. In analogy with a
continuous Carnot-like cycle, Dorfman $et$ $al.$ showed that the
power of a photocell based on Photosystem {\rm II} reaction center
(PS{\rm II} RC) can be increased by $27\% $ attributed to
noise-induced quantum coherence---Fano interference
\cite{Scully2011108,Svidzinsky201184}, which was found in artificial
photocells and lasing without inversion. Creatore $et$ $al.$
proposed  the other mechanism and showed that the dipole-dipole
interaction between two neighboring electron donors play a key role
in enhancing the current and power of the  photocell
\cite{Creatore2013111}. They claimed that this increasing  can be up
to $35\% $.

In these studies, only one charge-separation pathway is considered.
Identifying the primary electron donors and dominating
charge-separation pathways has been a question of recent extensive
research and debate. At the moment, there  is much evidence that two
main pathways make significant contribution under ambient condition
\cite{Novoderezhkin200793,Romero201049,Novoderezhkin201112,
Cardona20121817}. This motivates us to ask the question that whether
a multi-pathway scheme is more beneficial to designing artificial
light-harvesting devices? In this paper, we will answer this
question  via numerical simulation for the  current and power as
well as the robustness of the current and power against the charge
recombination and dephasing.

The remainder of the paper is organized as follows. In Sec. {\rm
II}, we introduce a model of a quantum heat engine inspired by
PS{\rm II} RC to describe the charge separation,    which includes a
second pathway in comparison with  the model proposed in
\cite{Dorfman2013110,Creatore2013111}. A master equation describing
the evolution  of the two-pathway system is also derived in this
section. In Sec. {\rm III}, the concepts of effective voltage $V$
and power are introduced and used to characterize  the charge
separation, new results due to
 the second pathway are discussed. In Sec. {\rm IV}, we focus on how the
multiple pathways affect the behavior of $j-V$ and $P-V$ features.
We explain the advantage of multiple pathways over simply increasing
the decay rates and cross-couplings. In Sec. {\rm V}, we take the
charge recombination and dephasing that might coexist in natural
PS{\rm II} RC into consideration. Numerical results of one- and
multi-pathway models are shown, from which we further see the
advantage of multiple pathways. Sec. {\rm VI} is devoted to
concluding remarks.
%%%%%%%%%%%%%%%%%%%%%%%%%%%%%%%%%%%%%%%%%%%%%%%%%%%%%%%%%
\section{Model}
\begin{figure*} \centering
\includegraphics[scale=0.20]{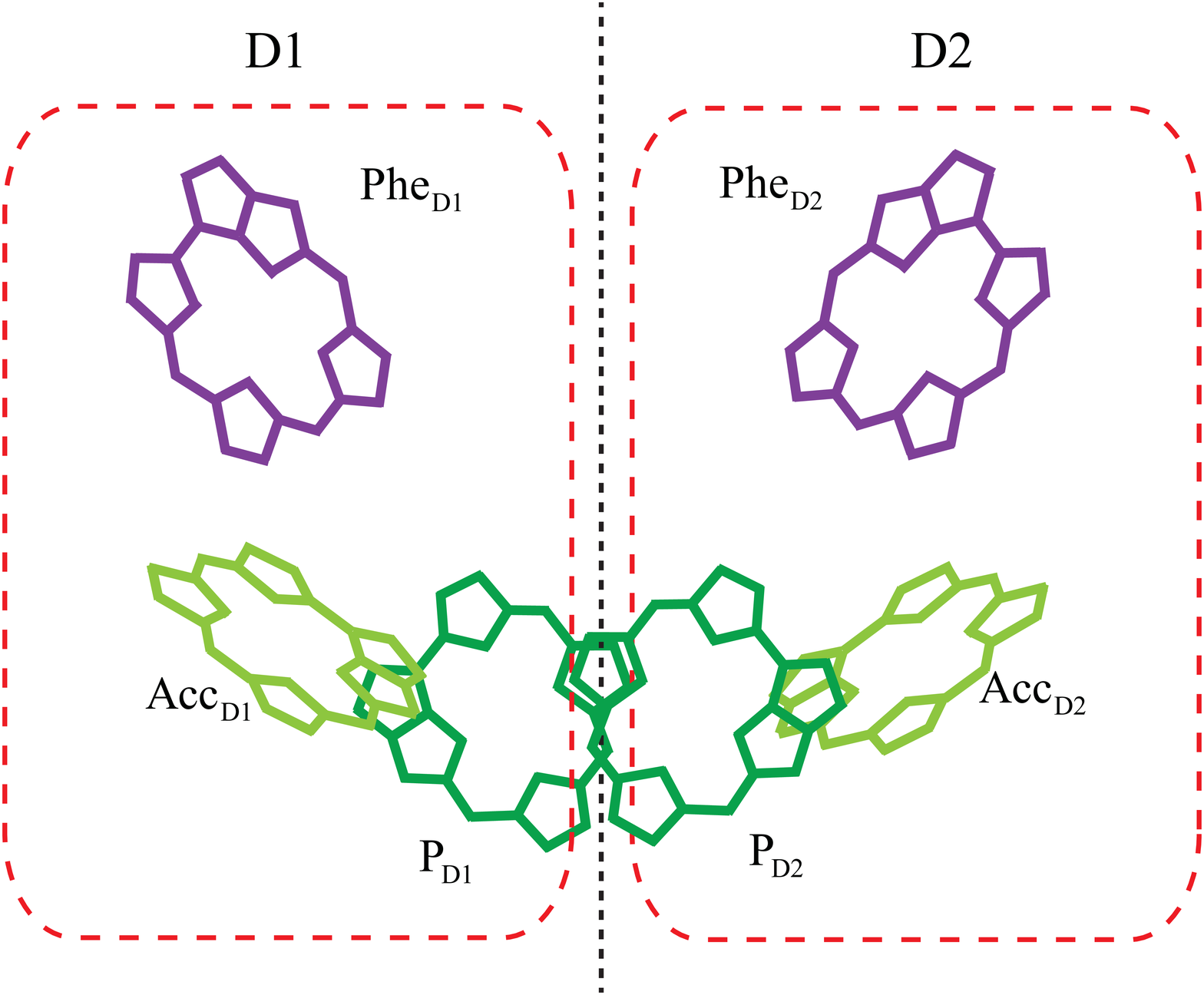}
\caption{(Color online) Arrangement of six core-pigments in the
PS{\rm II} RC. It consists of the special pair
${{\rm{P}}_{{\rm{D1}}}}$, ${{\rm{P}}_{{\rm{D2}}}}$, the two
accessory chlorophylls ${\rm{Ac}}{{\rm{c}}_{{\rm{D1}}}}$,
${\rm{Ac}}{{\rm{c}}_{{\rm{D2}}}}$, and the two pheophytins
${\rm{Ph}}{{\rm{e}}_{{\rm{D1}}}}$,
${\rm{Ph}}{{\rm{e}}_{{\rm{D2}}}}$.} \label{structureme:}
\end{figure*}

To present the model, we first illustrate  the structure of the
Photosystem II reaction center complex in Fig.~\ref{structureme:}.
The six pigment
\begin{figure*}
\centering
\includegraphics[scale=0.2]{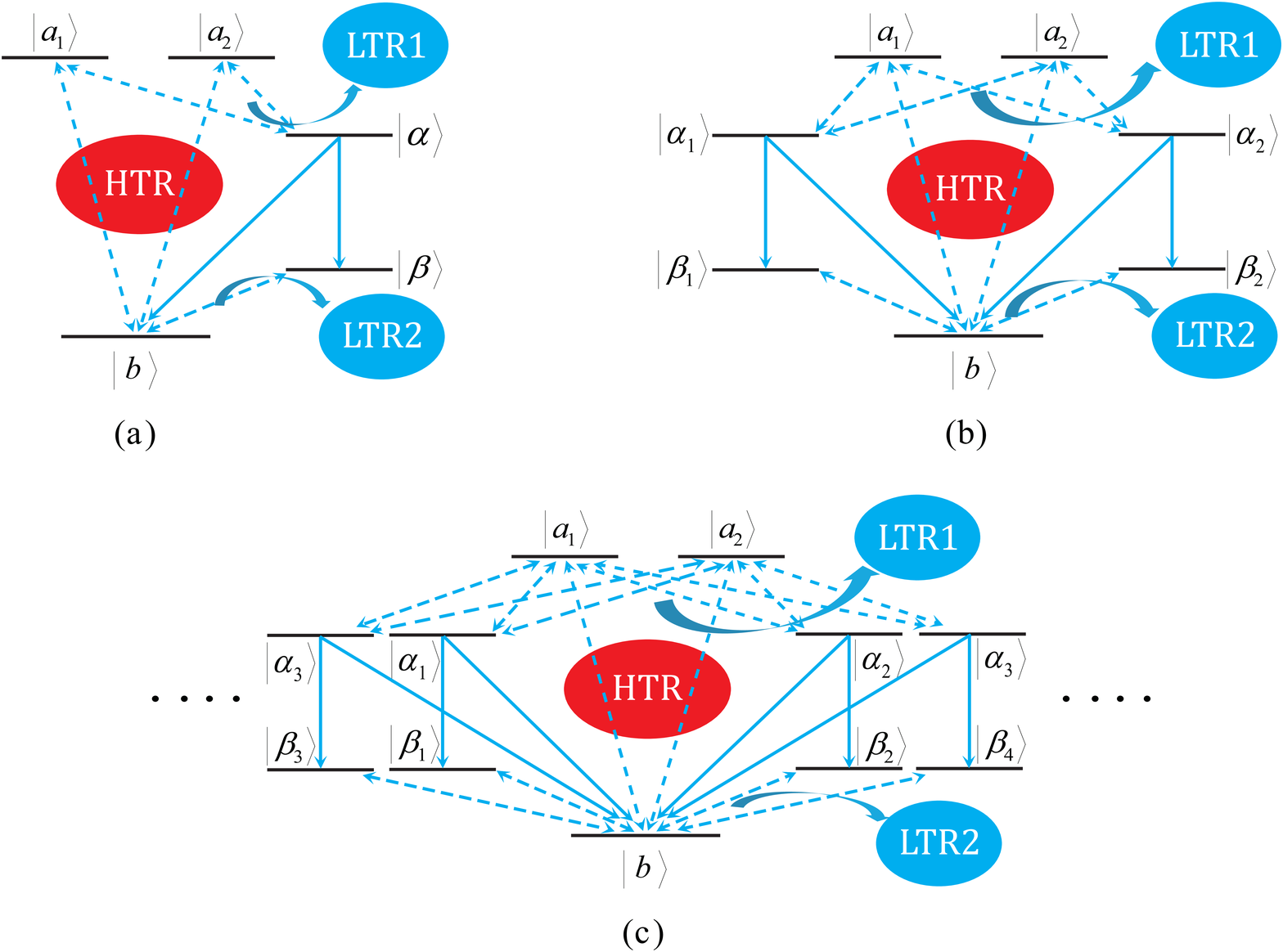}
\caption{(Color online) Schemes of the BQHE model based on the
photosynthetic reaction center. HTR denotes the high-temperature
photon reservoir, while LTR1 and LTR2 stand for the low-temperature
phonon reservoirs. HTR induces transition from the ground state
$\left| b \right\rangle $ to the single-exciton states $\left|
{{a_1}} \right\rangle $ and/or $\left| {{a_2}} \right\rangle $. LTR1
induces transition from $\left| {{a_1}} \right\rangle $ and/or
$\left| {{a_2}} \right\rangle $ to the charge-separated state
$\left| \alpha \right\rangle $ ($\left| {{\alpha _1}} \right\rangle
$ and $\left| {{\alpha _2}} \right\rangle $ for (b), $\left|
{{\alpha _i}} \right\rangle $ ($i = 1,2,3...$) for (c)).    LTR2
induces transition from the ionized state $\left| \beta
\right\rangle $ ($\left| {{\beta _1}} \right\rangle $ and $\left|
{{\beta _2}} \right\rangle $ for (b),
 $\left| {{\beta _i}} \right\rangle $ ($i = 1,2,3...$) for (c))
to the ground state. The three models differ in the number of donor,
i.e., pathway.}%Dorfman
\label{energystructure1:}
\end{figure*}
molecules are closely spaced in particular positions and
orientations, and coupled by the dipole-dipole interactions
resulting in  exciton states. These pigments are distributed in two
branches of protein matrix: $D1$ and $D2$. $P_{D1}$ and $P_{D2}$, a
special pair of coupled chlorophylls, is located at the center of
the PS{\rm II} RC. They contribute mostly to the lowest energy
states and are the primary electron donors, forming two exciton
states which are denoted as $\left| {{a_1}} \right\rangle $ and
$\left| {{a_2}} \right\rangle $. These two pairs of molecules are
also coupled to the accessory chlorophylls $Acc_{D1}$ and $Acc_{D2}$
located in the two different branches $D1$ and $D2$, respectively.
$Ph{e_{D1}}$ and $Ph{e_{D2}}$ are the remaining two pheophytin
pigments coupled to the rest of the molecules, and act as the
electron acceptors. Photosynthetic antennae transfer energy of the
absorbed solar photons to the RC where the transmembrane charge
separation takes place. Charge separation in the core of
pigment-protein  RC complexes is the first energy conversion step in
photosynthesis. The consequent electrochemical potential drives a
chain of chemical reactions, including the reduction of NADP to
NADPH, the synthesis of ATP and the oxidized part of the RC splits
water, releasing molecular oxygen, all these leading eventually to
the stable storage of solar energy.

The photosynthetic reaction center may be analyzed as a biological quantum heat engine%Dorfman
(BQHE) that transforms high-energy thermal photon radiation into
low-entropy electron flux. As shown in Fig.~\ref{energystructure1:}
(a), a five-level QHE scheme describes the photoinduced charge
separation between the donor and the acceptor molecules interacting
with thermal light. State $\left| b \right\rangle $ represents the
lowest energy configuration in which both donors and acceptor
molecules are in the ground state. States $\left| {{a_1}}
\right\rangle $ and $\left| {{a_2}} \right\rangle $ correspond to
single-exciton states in the first and second donors, respectively.
$\left| \alpha  \right\rangle $ is the charge-separated state with
the excited electron transferred from the donors to the acceptor
molecules. And $\left| \beta  \right\rangle $ describes the ionized
state in which the system is positively charged (the excited
electron is assumed to
have been used to perform work). %PRL
The heat engine cycle begins with the excited electron promoted from
$\left| b \right\rangle $ to $\left| {{a_1}} \right\rangle $ and/or
$\left| {{a_2}} \right\rangle $ after the absorption of a solar
photon. This excited electron is then transferred to the acceptor by
emission of a phonon (transition from $\left| {{a_1}} \right\rangle
$ and/or $\left| {{a_2}} \right\rangle $ to $\left| \alpha
\right\rangle $, corresponding to the formation of a radical pair
$P_{D2}^ + Acc_{D2}^ - $ in Fig.~\ref{structureme:}). Furthermore,
the electron can be released from state $\left| \alpha \right\rangle
$ with a rate $\Gamma $, resulting in a current from $\left| \alpha
\right\rangle $ to $\left| \beta  \right\rangle $ driving  a chain
of chemical reactions, leading eventually to the stable storage of
solar energy. The current $j = e\Gamma {\rho _{\alpha \alpha }}$ is
thus determined by the relaxation rate $\Gamma $ and the population
of $\left| \alpha  \right\rangle $. The acceptor-to-donor charge
recombination represented by the decay rate ${\Gamma _{\alpha \to
b}} = \chi \Gamma $, with $\chi $ a dimensionless fraction, brings
the system back to the ground state $\left| b \right\rangle $ but
does not produce current, limiting the power output of our QHE
\cite{Creatore2013111,Bredas200942}. Finally, to complete the cycle,
we allow another population transfer to take place, emitting a
phonon with excess energy, bringing the electron back to the neutral
ground state $\left| b \right\rangle $ with a rate ${\Gamma _c}$. We
also assume that the donor excited states $\left| {{a_1}}
\right\rangle $ and $\left| {{a_2}} \right\rangle $ undergo a
dephasing process (via a rate ${\Gamma _{dep}}$), which will
influence the current and power generated. This model including one
charge-separation pathway ($\left| \alpha \right\rangle  \to \left|
\beta  \right\rangle $) can exhibit noise-induced quantum coherence
due to Fano interference, which originates from the coupling of two
levels to a common reservoir and thus the effect can be revealed by
the cross-couplings. Fano interference can minimize the
acceptor-to-donor charge recombination by inducing coherence between
$\left| {{a_1}} \right\rangle $ and $\left| {{a_2}} \right\rangle $.
This brings about enhanced current and power output as we will show
in section II.

Although electron transfer in the PSII RC has been thoroughly
investigated and several charge-separation pathways that involve the
formation of several different radical pairs were identified, works
on the effect of multiple charge-separation pathways are rarely
found. At the moment, much evidence shows that two main pathways
make significant contributions under ambient conditions
\cite{Novoderezhkin200793,Romero201049,Novoderezhkin201112,
Cardona20121817}. Therefore we emphatically study two pathways,
i.e., we add another acceptor molecule that consists  of a
charged-separated state and an ionized state to the model in
Fig.~\ref{energystructure1:} (a). As shown in
Fig.~\ref{energystructure1:} (b), the initial excitation of states
$\left| {{a_1}} \right\rangle $ and $\left| {{a_2}} \right\rangle $
can be transferred to the two acceptor molecules in states $\left|
{{\alpha _1}} \right\rangle $ and $\left| {{\alpha _2}}
\right\rangle $, respectively, with the excess energy radiated as a
phonon, and further produce electric current and returning back to
$\left| b \right\rangle $ via $\left| {{\beta _1}} \right\rangle $
and/or $\left| {{\beta _2}} \right\rangle $. Here, the total current
in the two pathways is given by $j = e{\Gamma _1}{{ \rho } _{{\alpha
_1}{\alpha _1}}} + e{\Gamma _2}{{ \rho } _{{\alpha _2}{\alpha
_2}}}$. The structures of RCs from green plants, algae, and bacteria
differ in the nature and precise orientation of the constituent
pigments, resulting in different spectroscopic and dynamical
parameters. Nevertheless, our six-level scheme may be applied to all
RCs as discussed below.

With this knowledge, the dynamics of our heat engine can be
described by the following master equation,
\begin{eqnarray}
\frac{{d{{\hat \rho } _S}}}{{dt}} =  - i[{H_S},{{\hat \rho } _S}] +
L{{\hat \rho } _S}, \label{master equation}
\end{eqnarray}
where ${H_S} = \sum\limits_{\scriptstyle j = b,{a_1},{a_2}, \hfill
\atop \scriptstyle {\alpha _1},{\alpha _2},{\beta _1},{\beta _2}
\hfill}  {{E_j}\left| j \right\rangle \left\langle j \right|}  $ is
the system Hamiltonian, with ${{E_j}}$ the energy of the $i$th
level. The superoperator $L$ describing the effect of  reservoirs
and acceptor-donor recombinations can be decomposed as
\begin{eqnarray}
\begin{aligned}
L{{\hat \rho } _S} = & {L_h}{{\hat \rho } _S} +{L_{c1}}{{\hat \rho } _S}
+ {L_{c2}}{{\hat \rho } _S} + {L_{rel}}{{\hat \rho } _S}\\
& + {L_{dep}}{{\hat \rho } _S}+ {L_{rec}}{{\hat \rho } _S}.
\label{Lindblad}
\end{aligned}
\end{eqnarray}
For the high temperature photon
reservoir (HTR), its effects can be described by
\begin{eqnarray}
\begin{aligned}
{L_h}{{\hat \rho } _S} = &\sum\limits_{k,l = 1,2} {\frac{{{\gamma _{klh}}}}{2}}
 [({n_h} + 1)({\sigma _{hl}}{{\hat \rho } _S}\sigma _{hk}^\dag  + {\sigma _{hk}}{{\hat \rho }_S}\sigma _{hl}^\dag \\
&- \sigma _{hl}^\dag {\sigma _{hk}}{{\hat \rho } _S} -
{{\hat \rho } _S}\sigma _{hk}^\dag {\sigma _{hl}}) + {n_h}(\sigma _{hk}^\dag {{\hat \rho } _S}{\sigma _{hl}} \\
&+ \sigma _{hl}^\dag {{\hat \rho } _S}{\sigma _{hk}} - {\sigma
_{hk}}\sigma _{hl}^\dag {{\hat \rho } _S} - {{\hat \rho } _S}{\sigma
_{hl}}\sigma _{hk}^\dag )], \label{Lindblad h}
\end{aligned}
\end{eqnarray}
where ${{\hat \sigma }_{hk}} = \left| b \right\rangle \left\langle
{{a_{k}}} \right|$ $(k = 1,2)$, and  ${n_h}$ is the average number
of solar photons. ${\gamma _{klh}}$ ($k=l$) denotes the decay rate,
where ${\gamma _{11h}} = {\gamma _{1h}}$ and ${\gamma _{22h}} =
{\gamma _{2h}} $ represents the spontaneous decay from the upper
level $\left| {{a_1}} \right\rangle $ and $\left| {{a_2}}
\right\rangle $ to level $\left| b \right\rangle $, respectively.
The noise-induced quantum coherence is closely related to the terms
with  ${\gamma _{klh}}$ ($k \neq l $), which we  will refer to as
cross-couplings. They describe the effect of interference, and we
will assume ${\gamma _{12h}} = {\gamma _{21h}}$, and ${\gamma
_{klh}} = {\eta _h} \sqrt {{\gamma _{kh}}{\gamma _{lh}}} $ ($k \neq
l $) where ${\eta _h} =1$ represents  the maximal coherence and
${\eta _h} =0$ the minimal coherence. It is worth noting  that
${\eta _h} =0$ represents a case in which the two donor molecules
interact with their own (independent) reservoirs. Thus there is no
coherence induced by noise \cite{Xu201490} in this situation.

Superoperator ${L_{{c2}}}$ corresponding to the second low
temperature phonon reservoir (LTR2) has the form
\begin{eqnarray}
\begin{aligned}
{L_{c2}}{{\hat \rho } _S} = &\sum\limits_{k,l = 1,2}  {\frac{{{\Gamma _{klc}}}}{2}} [({n_{c2}} + 1)({\sigma _{cl}}{{\hat \rho } _S}\sigma _{ck}^\dag  + {\sigma _{ck}}{{\hat \rho } _S}\sigma _{cl}^\dag \\
 &- \sigma _{cl}^\dag {\sigma _{ck}}{{\hat \rho } _S} - {{\hat \rho } _S}\sigma _{ck}^\dag {\sigma _{cl}}) + {n_{c2}}(\sigma _{ck}^\dag {{\hat \rho } _S}{\sigma _{cl}} \\
 &+ \sigma _{cl}^\dag {{\hat \rho } _S}{\sigma _{ck}} - {\sigma _{ck}}\sigma _{cl}^\dag {{\hat \rho } _S} - {{\hat \rho } _S}{\sigma _{cl}}\sigma _{ck}^\dag )],
\label{Lindblad c2}
\end{aligned}
\end{eqnarray}
where ${{\hat \sigma }_{ck}} = \left| b \right\rangle \left\langle
{{\beta _{k}}} \right|$ $(k = 1,2)$ denotes the  lower operator.
${n_{c2}}$ is the average phonon number of the  cold (phonon)
reservoir LTR2. ${\Gamma _{klc}}$ ($k=l$) is the corresponding decay
rate, with ${\Gamma _{11c}} = {\Gamma _{1c}}$ and ${\Gamma _{22c}} =
{\Gamma _{2c}}$ representing the spontaneous decay from the upper
level $\left| {{\beta _1}} \right\rangle$ and $\left| {{\beta _2}}
\right\rangle $ to level $\left| b \right\rangle $, respectively;
and ${\Gamma _{klc}}$ ($k \neq l$) is a cross-coupling that
describes the effect of interference, with ${\Gamma _{12c}} =
{\Gamma _{21c}}$. In numerical simulations, we will set ${\Gamma
_{klc}} = {\eta _{{c_2}}} \sqrt {{\Gamma _{kc}}{\Gamma _{lc}}} $ ($k
\neq l$) with ${\eta _{{c_2}}}=1$ representing the maximal coherence
and ${\eta _{{c_2}}}=0$ the minimal coherence.

The quantum coherence induced by the low temperature phonon
reservoir (LTR1) can be described by,
\begin{eqnarray}
\begin{aligned}
{L_{c1}}{{\hat \rho } _S} =& \sum\limits_{k,l,m,n = 1,2} {\frac{{{\gamma _{klmn}}}}{2}} [({n_{c1}} + 1)({\sigma _{kl}}{{\hat \rho } _S}\sigma _{mn}^\dag \\
 &+ {\sigma _{mn}}{{\hat \rho } _S}\sigma _{kl}^\dag  - \sigma _{kl}^\dag {\sigma _{mn}}{{\hat \rho } _S} - {{\hat \rho } _S}\sigma _{mn}^\dag {\sigma _{kl}}) \\
 &+ {n_{c1}}(\sigma _{mn}^\dag {{\hat \rho } _S}{\sigma _{kl}} + \sigma _{kl}^\dag {{\hat \rho } _S}{\sigma _{mn}} - {\sigma _{mn}}\sigma _{kl}^\dag {{\hat \rho } _S}\\
 & - {{\hat \rho } _S}{\sigma _{kl}}\sigma _{mn}^\dag )],
\label{Lindblad c1}
\end{aligned}
\end{eqnarray}
where ${{\hat \sigma }_{kl}} = \left| {{\alpha _k}} \right\rangle
\left\langle {{a_l}} \right|$ $(k,l = 1,2)$ or ${{\hat \sigma
}_{mn}} = \left| {{\alpha _m}} \right\rangle \left\langle {{a_n}}
\right|(m,n = 1,2)$, and ${n_{c1}}$ is the average phonon numbers of
LTR1. ${\gamma _{klmn}}(kl = mn)$ are decay rates from the upper
levels $\left| {{a_1}} \right\rangle $ and $\left| {{a_2}}
\right\rangle $ to the lower levels $\left| {{\alpha _1}}
\right\rangle $ and $\left| {{\alpha _2}} \right\rangle $,
respectively. For $kl \ne mn$, terms with ${\gamma _{klmn}}$ would
induce quantum coherence   leading to   the  interference. Note that
the cross-couplings here are  complicated more than those of HTR or
LTR2. There are two types of cross-couplings due LTR1. One is the
same as  that in previous works \cite{Dorfman2013110,Xu201490},
defined as ${\gamma _{klmn}} =  \eta _{{c_1}}^1 \sqrt {{\gamma
_{kl}}{\gamma _{mn}}}$ ($kl \ne mn$), including ${\gamma _{1121}}$
and ${\gamma _{1222}}$. These two cross-couplings represented by
terms with ${\gamma _{12c}}$ in the model shown in
Fig.~\ref{energystructure1:} (a). The other one is defined as
${\gamma _{klmn}} =  \eta _{{c_1}}^2 \sqrt {{\gamma _{kl}}{\gamma
_{mn}}}$ ($kl \ne mn$), including terms with ${\gamma _{1112}}$,
${\gamma _{1122}}$, ${\gamma _{1221}}$, ${\gamma _{2122}}$. These
terms are not considered  in  the model depicted in
Fig.~\ref{energystructure1:} (a). The  first type of cross-couplings
couple $|a_1\rangle$ and $|a_2\rangle$ with  the same lower lever.
Namely, terms with ${\gamma _{1121}}$ couple $|a_1\rangle$ and
$|a_2\rangle$ to $\left| {{\alpha _1}} \right\rangle $, while terms
with ${\gamma _{1222}}$  to $\left| {{\alpha _2}} \right\rangle $.
In contrast, the second type of cross-couplings,  the lower levels
are different. In the later numerical simulations,  we will choose $
\eta _{{c_1}}^{1,2} = 1$ representing  the maximal coherence and $0$
the minimal coherence. We will discuss the effects caused by these
cross-couplings in section ${\rm{{\rm I}V}}$.

${L_{rel}}$ describes a  process that  the system in state $\left| {{\alpha
_1}} \right\rangle {\rm{(}}\left| {{\alpha _2}} \right\rangle
{\rm{)}}$ decays  to state $\left| {{\beta _1}} \right\rangle (\left|
{{\beta _2}} \right\rangle )$. It leads to the  electronic current
proportional to the relaxation rates ${\Gamma _1}$ and ${\Gamma _2}$
as defined later.
\begin{eqnarray}
\begin{aligned}
{L_{rel}}{{\hat \rho } _S} =& \sum\limits_{k = 1,2} {\frac{{{\Gamma _k}}}{2}} (\left| {{\beta _k}} \right\rangle \left\langle {{\alpha _k}} \right|{{\hat \rho } _S}\left| {{\alpha _k}} \right\rangle \left\langle {{\beta _k}} \right|\\
&- \left| {{\alpha _k}} \right\rangle \left\langle {{\alpha _k}}
\right|{{\hat \rho } _S} - {{\hat \rho } _S}\left| {{\alpha _k}}
\right\rangle \left\langle {{\alpha _k}} \right|). \label{Lindblad
relaxation}
\end{aligned}
\end{eqnarray}

We describe  the dephasing  of the system on states $\left| {{a_1}}
\right\rangle $ and $\left| {{a_2}} \right\rangle $ by
\begin{eqnarray}
\begin{aligned}
{L_{dep}}{{\hat \rho } _S} = &\sum\limits_{k = 1,2} {\frac{{{\Gamma
_{dep}}}}{2}}
(\left| {{a_k}} \right\rangle \left\langle {{a_k}} \right|{{\hat \rho }_S}\left| {{a_k}} \right\rangle \left\langle {{a_k}} \right| \\
&-\left| {{a_k}} \right\rangle \left\langle {{a_k}} \right|{{\hat
\rho } _S} - {{\hat \rho } _S}\left| {{a_k}} \right\rangle
\left\langle {{a_k}} \right|). \label{Lindblad decoherence}
\end{aligned}
\end{eqnarray}
with ${{\Gamma _{dep}}}$ being the dephasing rate.

Finally, the loss channel due to acceptor-to-donor charge  recombination
is described by,
\begin{eqnarray}
\begin{aligned}
{L_{rec}}{{\hat \rho } _S} = &\sum\limits_{k = 1,2}  {\frac{{\chi {\Gamma _k}}}{2}} (\left| b \right\rangle \left\langle {{\alpha _k}} \right|{{\hat \rho } _S}\left| {{\alpha _k}} \right\rangle \left\langle b \right| \\
&- \left| {{\alpha _k}} \right\rangle \left\langle {{\alpha _k}}
\right|{{\hat \rho } _S} - {{\hat \rho } _S}\left| {{\alpha _k}}
\right\rangle \left\langle {{\alpha _k}} \right|). \label{Lindblad
recombination}
\end{aligned}
\end{eqnarray}
Here the dimensionless $\chi$ stands for the recombination rate.

\section{Effects of cross-couplings}
%table 1
\begin{table}[h]
\centering \caption{Parameters used in the numerical simulations.}
\includegraphics[scale=0.426]{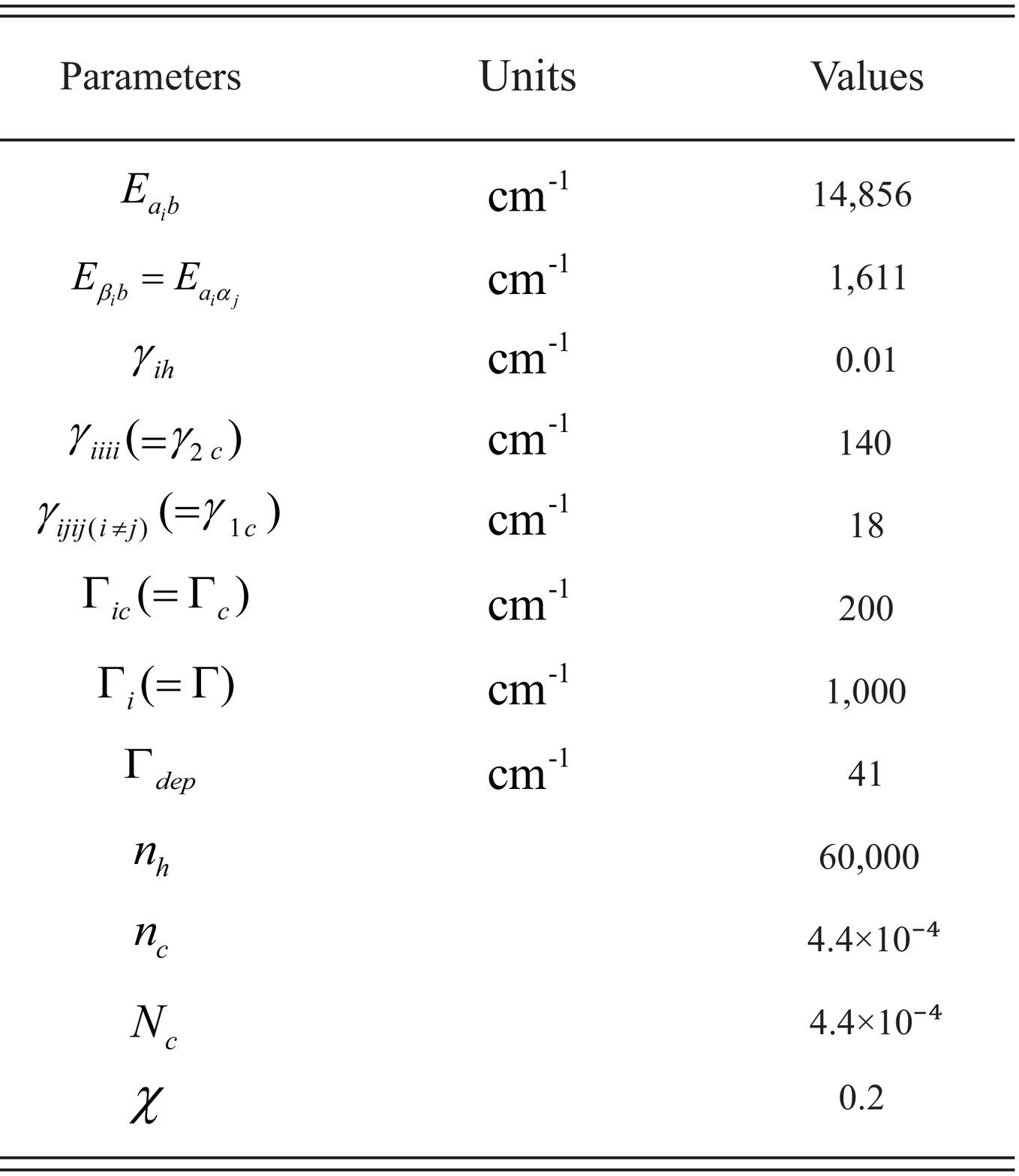}
\label{table:}
\end{table}
%PRL
In the present  system, the current generated can be thought to flow
across a load connecting the acceptor levels $\left| \alpha
\right\rangle $ and $\left| \beta  \right\rangle $. Introducing an
effective voltage $V$ as a drop of the electrostatic
potential across the load,  we obtain $eV = {E_\alpha } - {E_\beta }
+ {k_B}T\ln (\frac{{{{\hat \rho } _{\alpha \alpha }}}}{{{{\hat \rho
} _{\beta \beta }}}})$ for the model in Fig.~\ref{energystructure1:}
(a), where $e$ is the electric charge and ${E_i}$ is the energy of
state $\left| i \right\rangle $.

In our model  shown in Fig.~\ref{energystructure1:} (b), choosing
${E_{{i_1}}} = {E_{{i_2}}}$ ($i = a, \alpha ,\beta $) and setting
all parameters the same for the two pathways, we have $eV =
{E_{{\alpha _1}}} - {E_{{\beta _1}}} + {k_B}T\ln (\frac{{{{ \rho }
_{{\alpha _1}{\alpha _1}}}}}{{{{ \rho } _{{\beta _1}{\beta
_1}}}}}),$ and $eV = {E_{{\alpha _2}}} - {E_{{\beta _2}}} +
{k_B}T\ln (\frac{{{{ \rho } _{{\alpha _2}{\alpha _2}}}}}{{{{ \rho }
_{{\beta _2}{\beta _2}}}}})$, which define the   voltage $V$. The
current, accordingly, is defined  by $j = e{\Gamma _1}{{ \rho }
_{{\alpha _1}{\alpha _1}}} + e{\Gamma _2}{{ \rho } _{{\alpha
_2}{\alpha _2}}}$. It is similar to the parallel circuit in
classical electromagnetism. Based on the current and the voltage, we
easily obtain the power output $P = j \cdot V$. We apply this to the
steady-state of the system to characterize  the performance of our
BQHE  and the  photovoltaic properties of the complex, i.e., the
steady-state current-voltage ($j - V$) and power-voltage ($P - V$)
characteristics. Using the steady-state solution of Eq. (\ref{master
equation}), we plot the $j - V$  curve and power with different
$\Gamma $, while keep the other parameters fixed. Note that  $\Gamma
\to 0$ ($j \to 0$) corresponds  to the open-circuit case, and in the
short-circuit case,  $V \to 0$.

The parameters used in our simulation are listed in Table ${\rm I}$.
These   parameters  are chosen from recent literature, e.g.,
\cite{Dorfman2013110,Abramavicius2010133185401,Madjet2006110,Abramavicius2010133}
and they are used in the simulation in
\cite{Creatore2013111,Dorfman2013110,Xu201490}. The energy
differences are defined as ${E_{{a_i}b}} = {E_{{a_i}}} - {E_b}$,
${E_{{\beta _i}b}} = {E_{{\beta _i}}} - {E_b}$ and ${E_{{a_i}{\alpha
_j}}} = {E_{{a_i}}} - {E_{{\alpha _j}}}$, where $i,j=1,2$.
Parameters  in the brackets  of Table  ${\rm I}$ are used in the
one-pathway model in Fig.~\ref{energystructure1:} (a). In the
following, we will discuss the effect of cross-couplings separately.

\subsection{Cross-coupling ${\Gamma _{12c}}$ induced by LTR2}
%Fig1
\begin{figure}[htbp]
\centering
\includegraphics[scale=0.225]{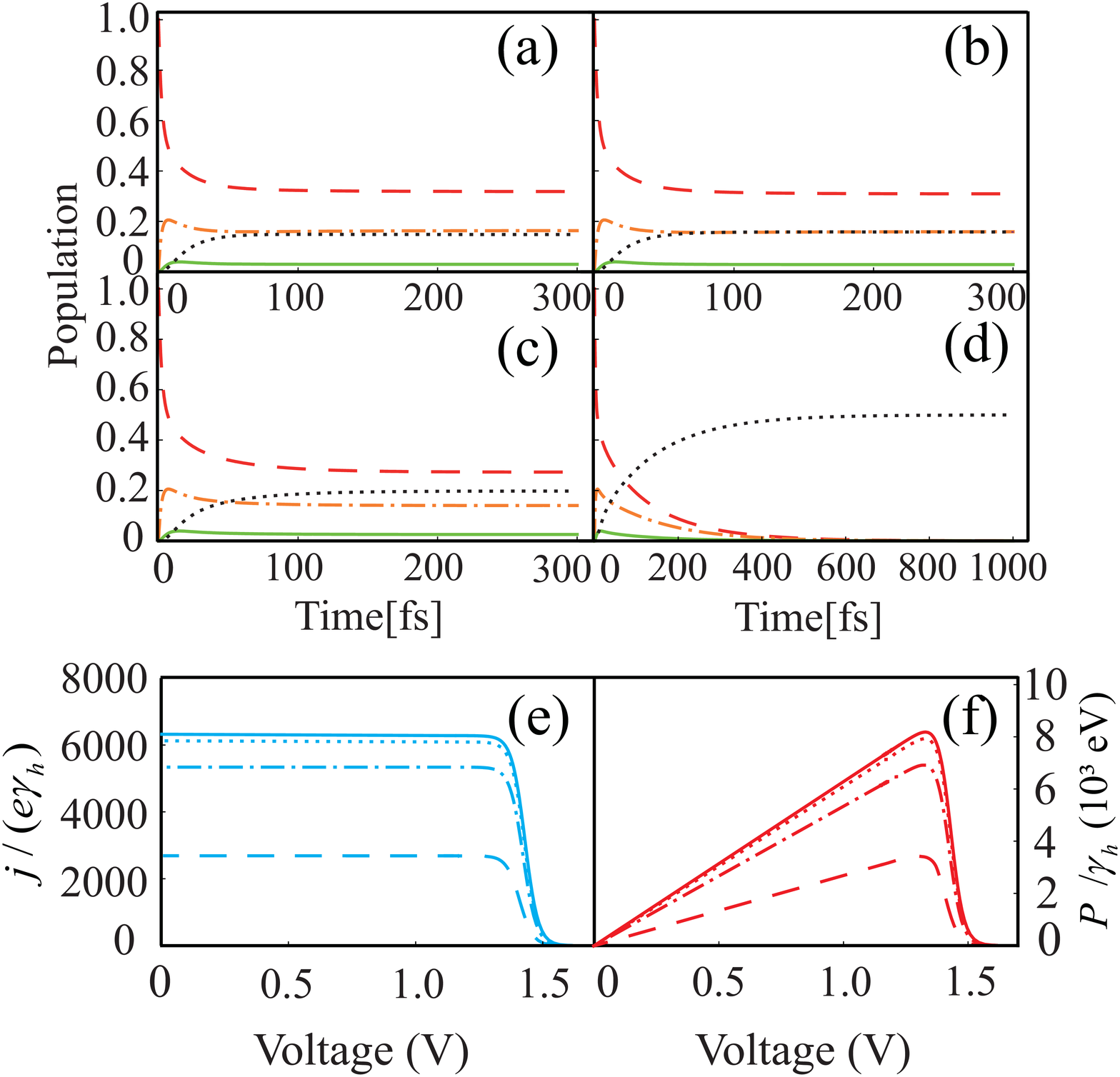}
\caption{(Color online) Fig.~\ref{Gamma12c:} (a)-(d) show the time
evolution of the population on each level of two-pathway model
corresponding to Fig.~\ref{energystructure1:} (b) at room
temperature $T = 300{\rm{K}}$: ${\rho _{bb}}$ in red-dashed, ${\rho
_{{a_1}{a_1}}}$ (${\rho _{{a_2}{a_2}}}$) in orange-dot-dashed,
${\rho _{{\alpha _1}{\alpha _1}}}$ (${\rho _{{\alpha _2}{\alpha
_2}}}$) in green-solid, and ${\rho _{{\beta _1}{\beta _1}}}$ (${\rho
_{{\beta _2}{\beta _2}}}$) in black-dotted line. The cross-coupling
${\Gamma _{12c}} ={\eta _{{c_2}}}\sqrt {{\Gamma _{1c}}{\Gamma
_{2c}}} $ takes ${\eta _{{c_2}}} = 0, 0.3, 0.6, 1$ for (a), (b), (c)
and (d), respectively. The other cross-couplings ${\gamma _{12h}} =
{\eta _{h}}\sqrt {{\gamma _{1h}}{\gamma _{2h}}} $, ${\gamma _{klmn}}
= \eta _{{c_1}}^1
 \sqrt {{\gamma _{kl}}{\gamma _{mn}}} $ ($kl \neq mn$) and ${\gamma _{klmn}} =
\eta _{{c_1}}^2 \sqrt {{\gamma _{kl}}{\gamma _{mn}}} $ ($kl \neq
mn$) take the maximal values, i.e., ${\eta _h} = \eta _{{c_1}}^1 =
\eta _{{c_1}}^2 = 1$. (e) and (f) are plotted for the current and
power generated as a function of effective voltage $V$ at room
temperature: ${\eta _{{c_2}}}= 0$ in solid, ${\eta _{{c_2}}} = 0.3$
in dotted, ${\eta _{{c_2}}} = 0.6$ \textbf{in} dot-dashed and ${\eta
_{{c_2}}}=
0.9$ in dashed line. }%Dorfman
\label{Gamma12c:}
\end{figure}
Interestingly, the $j - V$ and $P - V$ behaviors demonstrate that as
cross-coupling strength ${\Gamma _{12c}}$ increases, the current and
power decrease, see Fig.~\ref{Gamma12c:} (e) and (f). This is
different from the results in  previous works
\cite{Dorfman2013110,Xu201490}. Note that, the parameters chosen for
the two pathways are the same (see, Table ${\rm I}$), thus the
population on $\left| {{a_1}} \right\rangle $ and $\left| {{a_2}}
\right\rangle $ is the same, so is the population on $\left|
{{\alpha _1}} \right\rangle $ and $\left| {{\alpha _2}}
\right\rangle $, $\left| {{\beta _1}} \right\rangle $ and $\left|
{{\beta _2}} \right\rangle $. From Fig.~\ref{Gamma12c:} (d), we see
that, the maximal ${\Gamma _{12c}}$ (${\eta _{{c_2}}}=1$) benefits
the population on $\left| {{\beta _1}} \right\rangle $ and $\left|
{{\beta _2}} \right\rangle $ and leads this population dominant over
the others. Therefore we cannot plot   $j-V$ characteristics in this
case as ${\rho _{{\alpha _1}{\alpha _1}}} = {\rho _{{\alpha
_2}{\alpha _2}}} = 0$. It was shown  in Ref.
\cite{Dorfman2013110,Xu201490} that the power is increased and the
current-voltage characteristic of the heat engine gets better with
Fano interference. Here Fano interference characterized by ${\Gamma
_{12c}}$, however, shows an opposite  effect. We will explain this
observation  in subsection D. So, in the following numerical
simulations, we will take ${\Gamma _{12c}}=0$.

\subsection{Cross-couplings induced by LTR1}

\begin{figure}[htbp]
\centering
\includegraphics[scale=0.225]{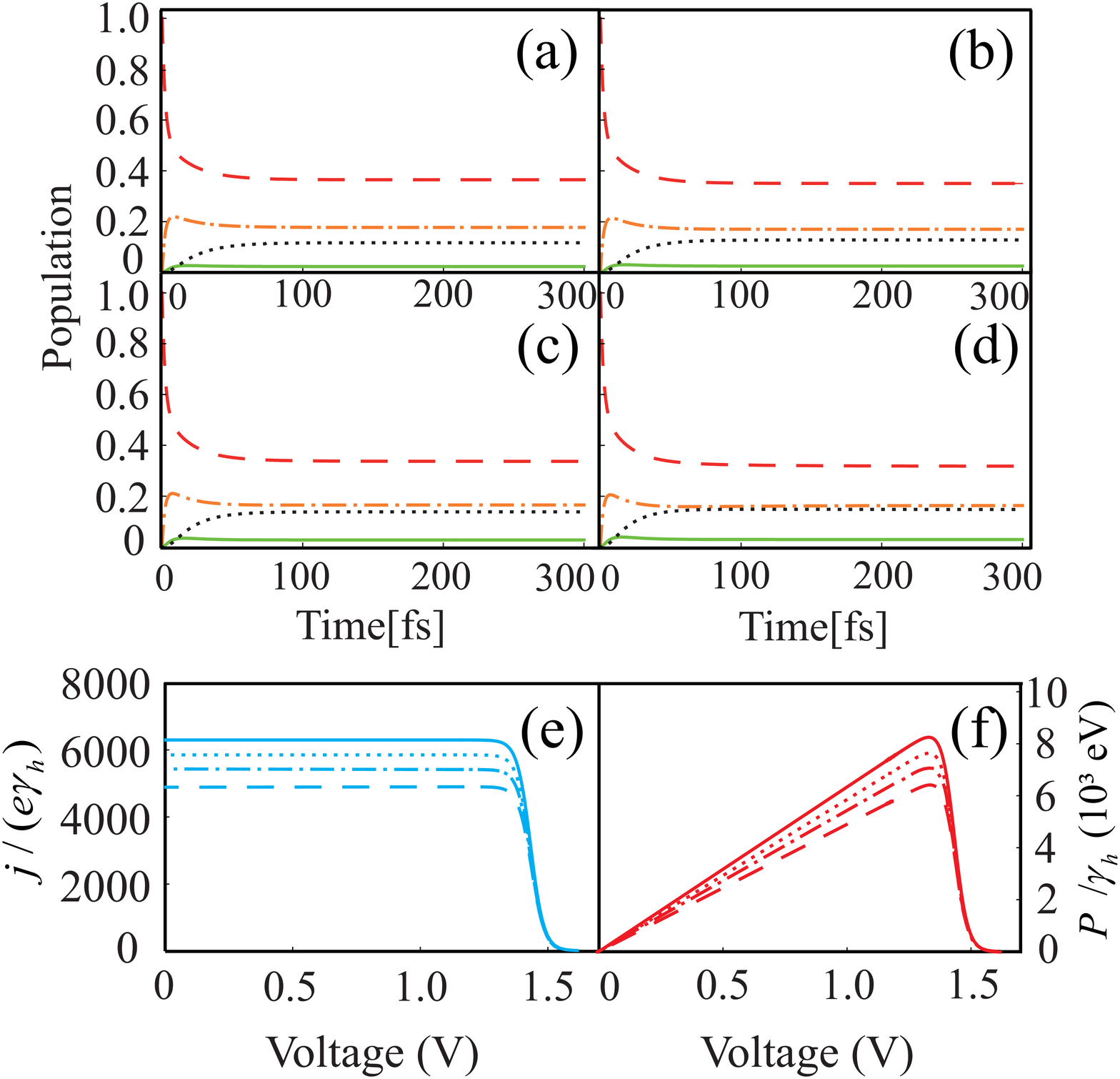}
\caption{(Color online) Fig.~\ref{gamma11:} (a) - (d) show the time
evolution of the population on each level in the two-pathway model
shown in  Fig.~\ref{energystructure1:} (b) at room temperature $T =
300{\rm{K}}$. ${\rho _{bb}}$ was plotted in red-dashed, ${\rho
_{{a_1}{a_1}}}$ (${\rho _{{a_2}{a_2}}}$) in orange-dot-dashed,
${\rho _{{\alpha _1}{\alpha _1}}}$ (${\rho _{{\alpha _2}{\alpha
_2}}}$) in green-solid, and ${\rho _{{\beta _1}{\beta _1}}}$ (${\rho
_{{\beta _2}{\beta _2}}}$) in black-dotted line. $\eta _{{c_1}}^1$
in the cross-couplings ${\gamma _{klmn}} = \eta _{{c_1}}^1\sqrt
{{\gamma _{kl}}{\gamma _{mn}}} $ ($kl \neq mn$) takes $\eta
_{{c_1}}^1 = 0, 0.3, 0.6, 1$ for (a), (b), (c) and (d),
respectively. The other cross-couplings ${\gamma _{12h}} = \eta
_{{h}} \sqrt {{\gamma _{1h}}{\gamma _{2h}}} $ and ${\gamma _{klmn}}
= \eta _{{c_1}}^2\sqrt {{\gamma _{kl}}{\gamma _{mn}}} $ ($kl \neq
mn$) take the maximal values, i.e., ${\eta _h} = \eta _{{c_1}}^2 =
1$. As mentioned in the last subsection, ${\Gamma _{12c}} = 0$. (e)
and (f) are plotted for the current and power generated as a
function of effective voltage $V$ at room temperature.  Dashed line
is for $\eta_{{c_1}}^1= 0$, dot-dashed for $\eta _{{c_1}}^1 = 0.3$,
dotted line for $\eta_{{c_1}}^1 = 0.6$, and $\eta _{{c_1}}^1 = 1$ is
ploted in
thin-solid line.}%Dorfman
\label{gamma11:}
\end{figure}

Fig.~\ref{gamma11:} shows that, current and power are  enhanced as
 ${\gamma _{11,21}}$ and ${\gamma _{12,22}}$ increase, which coincides with
the results predicted by Dorfman $et$ $al$. This gives rise to a
question that  why does ${\Gamma _{12c}}$ exhibit effects different
from that by ${\gamma _{11,21}}$ and ${\gamma _{12,22}}$?

%Fig1

\begin{figure}[htbp]
\centering
\includegraphics[scale=0.225]{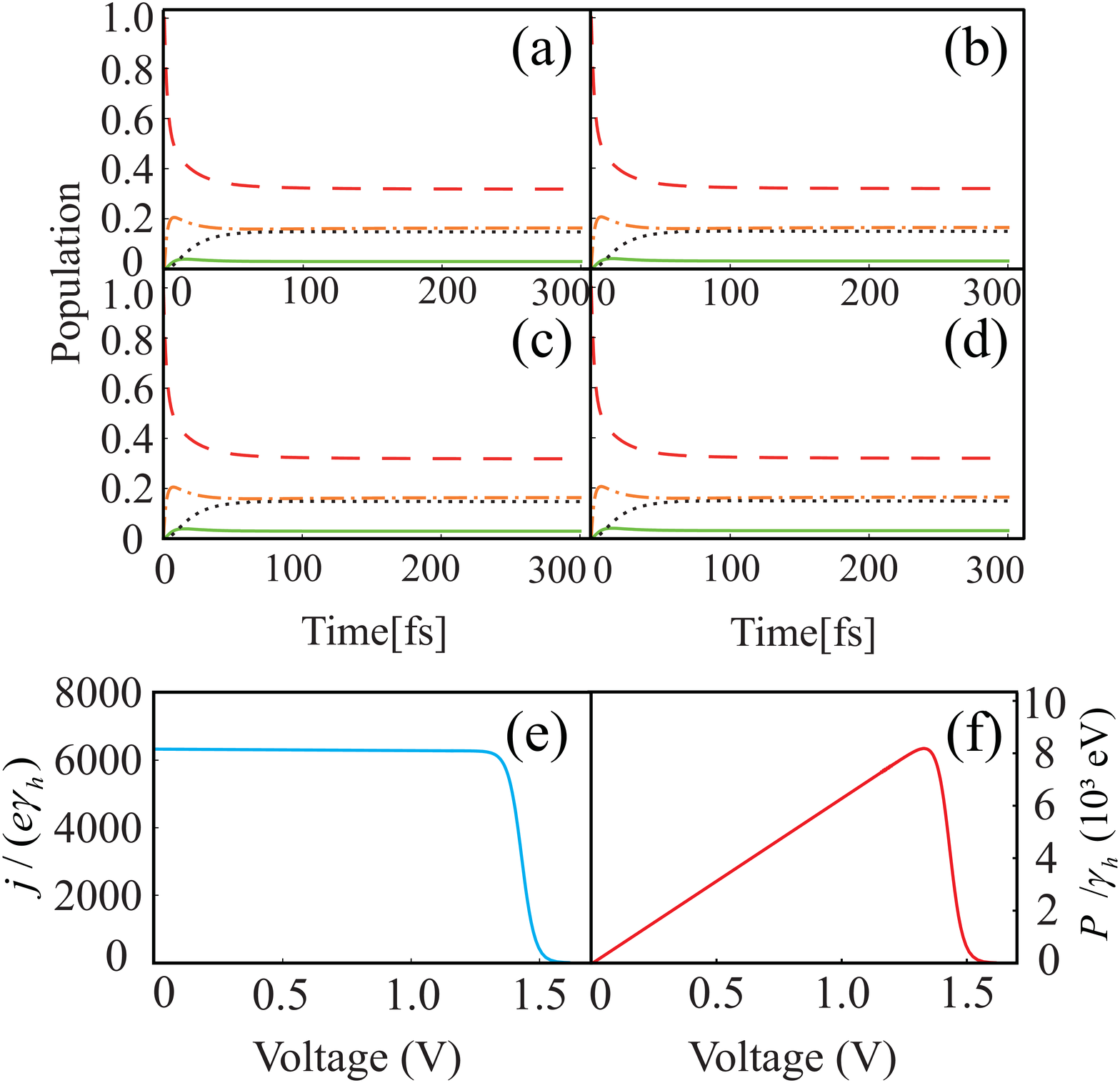}
\caption{(Color online) Fig.~\ref{gamma12:} (a)-(d) show the time
evolution of the population on each level of the system depicted in
Fig.~\ref{energystructure1:} (b) at room temperature $T =
300{\rm{K}}$: ${\rho _{bb}}$ in red-dashed, ${\rho _{{a_1}{a_1}}}$
(${\rho _{{a_2}{a_2}}}$) in orange-dot-dashed, ${\rho _{{\alpha
_1}{\alpha _1}}}$ (${\rho _{{\alpha _2}{\alpha _2}}}$) in
green-solid, and ${\rho _{{\beta _1}{\beta _1}}}$ (${\rho _{{\beta
_2}{\beta _2}}}$) in black-dotted line. The cross-couplings ${\gamma
_{klmn}} = \eta _{{c_1}}^2 \sqrt {{\gamma _{kl}}{\gamma _{mn}}} $
($kl \neq mn$) take $\eta _{{c_1}}^2 = 0, 0.3, 0.6, 1$ for (a), (b),
(c) and (d), respectively. The other cross-couplings ${\gamma
_{12h}} = \eta _{h}\sqrt {{\gamma _{1h}}{\gamma _{2h}}} $ and
${\gamma _{klmn}} = \eta _{{c_1}}^1 \sqrt {{\gamma _{kl}}{\gamma
_{mn}}} $ ($kl \neq mn$) take the maximal values, i.e., ${\eta _h} =
\eta _{{c_1}}^1  = 1$.  ${\Gamma _{12c}} = 0$. (e) and (f) are
plotted for the current and power as a function of effective voltage
$V$ at room temperature $T = 300{\rm{K}}$. Actually, we plot four
curves in (e) and (f) with $\eta _{{c_1}}^2 = 0, 0.3, 0.6, 1$,
respectively, although all  them  overlap. Since (a)-(d) do not show
any difference for different  $\eta _{{c_1}}^2 $,
the $j-V$ and $P-V$ behaviors are exactly the same for different $\eta _{{c_1}}^2 $.}%Dorfman
\label{gamma12:}
\end{figure}

\begin{figure}[htbp]
\centering
\includegraphics[scale=0.22]{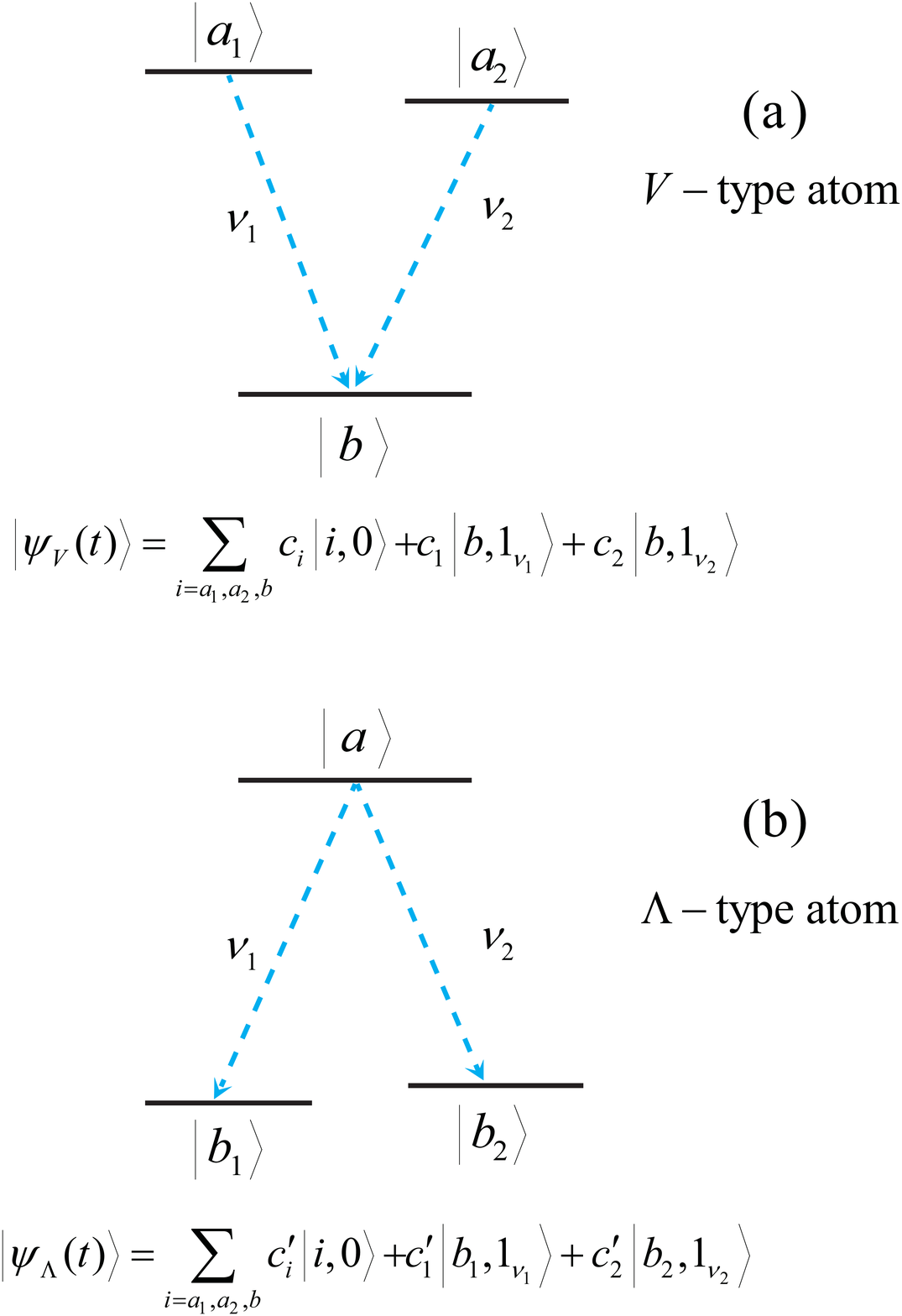}
\caption{(Color online) Three-level atomic structures for (a)
$V$-type and (b) $\Lambda$-type quantum beat.} \label{quantumbeat:}
\end{figure}

Fig.~\ref{gamma12:} shows that, population in each state seems no
difference for any values of the second type of cross-couplings
${\gamma _{klmn}} = \eta _{{c_1}}^2 \sqrt {{\gamma _{kl}}{\gamma
_{mn}}}$ ($kl \neq mn$). Thus different from couplings  with
${\gamma _{1121}}$ (${\gamma _{1222}}$) and ${\gamma _{12c}}$, the
second type of cross-couplings has marginal effects  on the current
and  power. This is analogous to quantum beats, which can occur in a
$V$ type atom, but can not in  a $\Lambda $ type atom based on the
theory of  quantum electrodynamics (QED)\cite{Scully1997}, as shown
in Fig.~\ref{quantumbeat:}.
%Fig1
$V$-type atom in excited states decays via the emission of a photon
with frequency ${\nu _1}$ or ${\nu _2}$. Since both emissions lead
to the same final state, we cannot determine along $which$ $path$,
${\nu _1}$ or ${\nu _2}$, the atom decays. This uncertainty   leads
to an interference between ${\nu _1}$ and ${\nu _2}$, exhibiting
quantum beats. A $\Lambda$-type atom also decays via the emission of
a photon with frequency ${\nu_1}$ or ${\nu _2}$. However, an
observation of atom at a long time after the emission can tell us
which channel (1 or 2) was taken (atom in $\left| {{b_1}}
\right\rangle $ or $\left| {{b_2}} \right\rangle $). Consequently,
we expect no beats in this case. For the model of
Fig.~\ref{energystructure1:} (b), transitions  described by the
second type of cross-couplings lead to different lower states, which
do not induce interference. Thus the second type of cross-couplings
has no effect on the $j - V$ and  $P - V$ behaviors.

With this consideration, in the following numerical  simulations, we
take ${\gamma_{klmn}} = \eta _{{c_1}}^2 \sqrt {{\gamma _{kl}}{\gamma
_{mn}}} =0$.

\subsection{Cross-couplings induced by HTR}

\begin{figure}[htbp]
\centering
\includegraphics[scale=0.22]{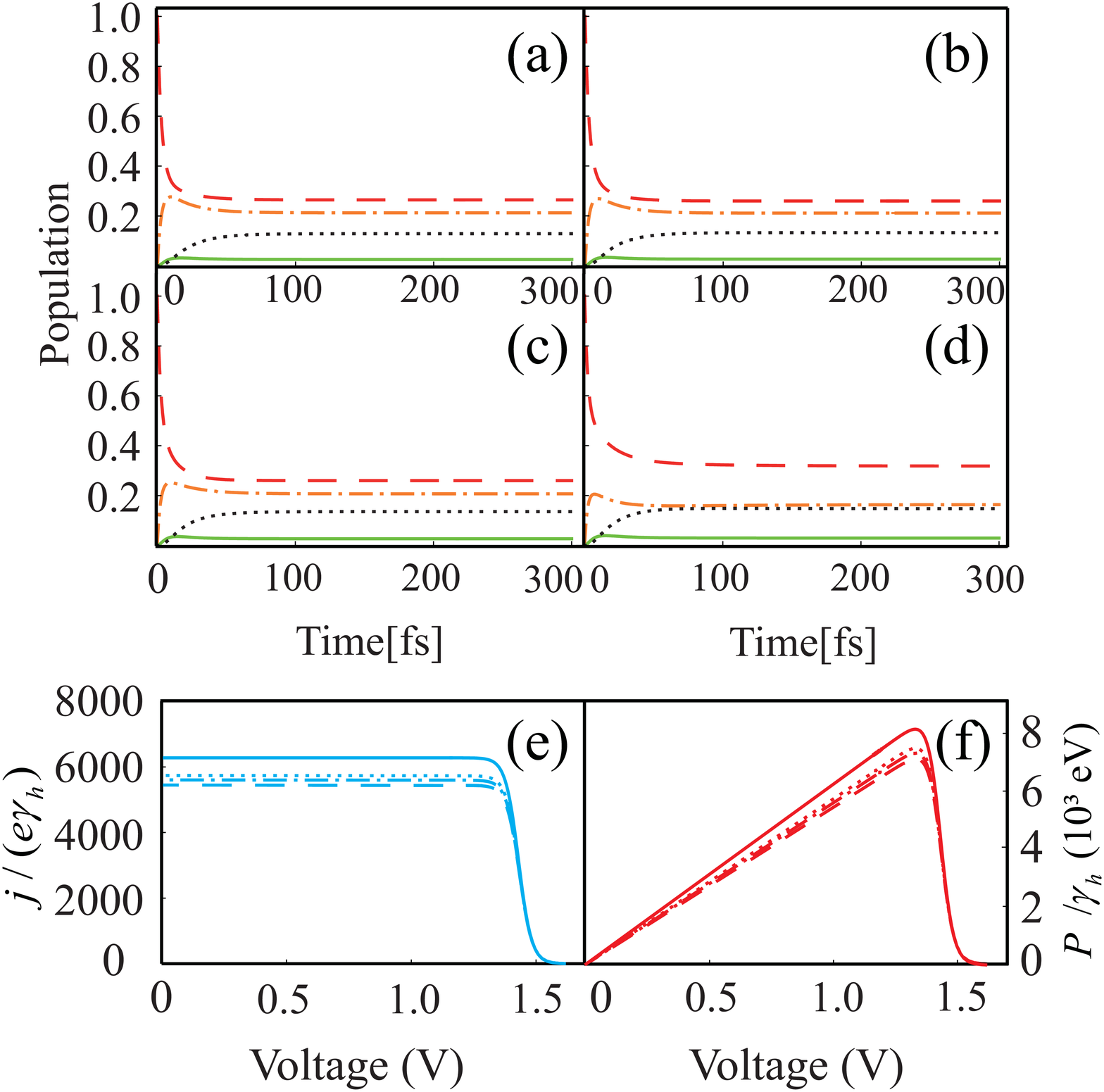}
\caption{(Color online) Fig.~\ref{gammah:} (a)-(d) show the time
evolution of the population on each level of the system in Fig.~\ref{energystructure1:} (b) at room
temperature $T = 300{\rm{K}}$. ${\rho _{bb}}$ is shown in red-dashed, ${\rho
_{{a_1}{a_1}}}$ (${\rho _{{a_2}{a_2}}}$) in orange-dot-dashed,
${\rho _{{\alpha _1}{\alpha _1}}}$ (${\rho _{{\alpha _2}{\alpha
_2}}}$) in green-solid, and ${\rho _{{\beta _1}{\beta _1}}}$ (${\rho
_{{\beta _2}{\beta _2}}}$) in black-dotted line. The cross-couplings
${\gamma _{12h}} = \eta _{h} \sqrt {{\gamma _{1h}}{\gamma _{2h}}} $
take $\eta _{h} = 0, 0.3, 0.6, 1$ for (a), (b), (c) and (d),
respectively.  The other cross-couplings ${\gamma _{klmn}} = \eta
_{{c_1}}^1 \sqrt {{\gamma _{kl}}{\gamma _{mn}}} $ ($kl \neq mn$)
takes the maximal value, i.e., $\eta _{{c_1}}^1 = 1$. ${\gamma
_{klmn}} = \eta _{{c_1}}^2 \sqrt {{\gamma _{kl}}{\gamma _{mn}}} = 0$
($kl \neq mn$) and ${\Gamma _{12c}} = 0$. (e) and (f) are plotted
for the current and power  as a function of effective
voltage $V$ at room temperature. $\eta _{h}= 0$ is plotted in dashed, $\eta
_{h} = 0.3$ in dot-dashed, $\eta _{h} = 0.6$ in dotted and $\eta _{h} =
1$ in
thin-solid line.}
\label{gammah:}
\end{figure}

As expected, ${\gamma _{12h}}$ benefits  the transition as seen from
Fig.~\ref{gammah:}. Inspired by the physics behind the quantum beat,
we claim that interference can play an important role in the current
and power only when the upper two levels $\left| {{a_1}}
\right\rangle $ and $\left| {{a_2}} \right\rangle $  are coherently
excited at the initial time. When the  coherence between $\left|
{{a_1}} \right\rangle $ and $\left| {{a_2}} \right\rangle $ is
absent, i.e., seting ${\gamma _{12h}} = 0$, ${\gamma _{12c}}$ itself
may suppress the exciton transfer. Equivalently,  setting ${\gamma
_{12c}} = 0$, ${\gamma _{12h}}$ itself may inhibit the current and
power. In terms of steady-state solution to Eq. \ref{master
equation}, we further confirm this observation, and it shows how
these two types of interference work together to affect the
transitions.

%Fig1

\subsection{Combined effects of HTR and LTR1}

\begin{figure}[htbp]
\centering
\includegraphics[scale=0.3]{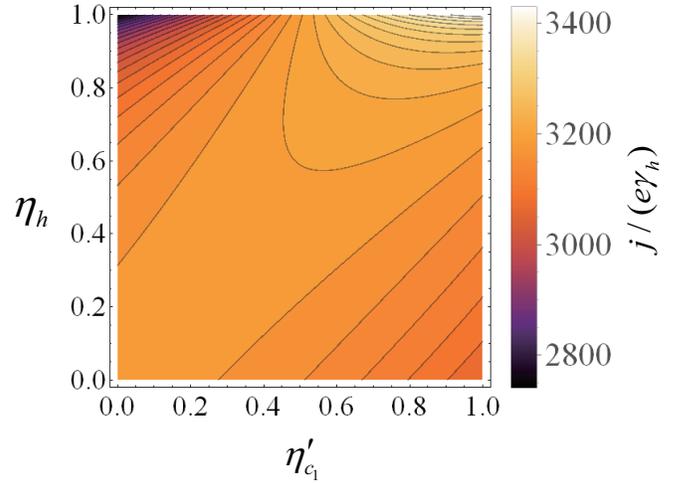}
\caption{(Color online) Counter plot of the current as a function of
cross-couplings ${\gamma _{12c}}$ and ${\gamma _{12h}}$. As ${\gamma
_{12c}} = {{\eta '}_{{c_1}}} \sqrt {{\gamma _{1c}}{\gamma _{2c}}} $
$({\gamma _{12h}} = {\eta _{h}}\sqrt {{\gamma _{1h}}{\gamma _{2h}}}
)$, $ {{\eta '}_{{c_1}}} = 1$ ${\rm{({\eta _h} = 1)}}$ corresponds
to the maximal coherence and ${{\eta '}_{{c_1}}}
= 0$ ${\rm{({\eta _{h}} = 0)}}$ the minimal coherence.}%Dorfman
\label{combined:}
\end{figure}

For simplicity, we plot the steady-state $j-V$ characteristic and
power   for one-pathway model in Fig.~\ref{combined:}. We set ${\gamma _{12c}} =
{\eta' _{{c_1}}} \sqrt {{\gamma _{1c}}{\gamma _{2c}}}  $ and
${\gamma _{12h}} = {\eta _{h}} \sqrt {{\gamma _{1h}}{\gamma _{2h}}}
$, thus ${\eta' _{{c_1}}} = 1$ (${\eta _{h}} = 1$) corresponds to
the maximal coherence and ${\eta' _{{c_1}}} = 0$ (${\eta _{h}} = 0$)
the minimal coherence.
%Fig1
 We observe that, when
${\gamma _{12h}}$ is very small, the current and power decrease with
the increasing of  ${\gamma _{12c}}$. Similarly, when ${\gamma
_{12c}}$ is very small, large  ${\gamma _{12h}}$ inhibits the
transitions. Recall the  quantum beats, if one of ${\gamma _{12h}}$
and ${\gamma _{12c}}$ is very small, $\left| {{a_1}} \right\rangle $
and $\left| {{a_2}} \right\rangle $ are almost independently
excited. For this reason, when one cross-coupling term becomes very
large,  the current and power will be suppressed. When both ${\gamma
_{12h}}$ and ${\gamma _{12c}}$ increase to a certain value, strong
cross-couplings can enhance the current and power. This conclusion
holds for ${\gamma _{1121}}$ and ${\gamma _{1222}}$ in the
two-pathway model. For the terms with ${\Gamma _{12c}}$ discussed in
subsection A, $\left| {{\beta _1}} \right\rangle $ and $\left|
{{\beta _2}} \right\rangle $ obviously, are not in a coherent state.
As a result, the transition process is suppressed as ${\Gamma
_{12c}}$ increases.

\section{Effect of Multiple pathways}

\begin{figure}[htbp]
\centering
\includegraphics[scale=0.26]{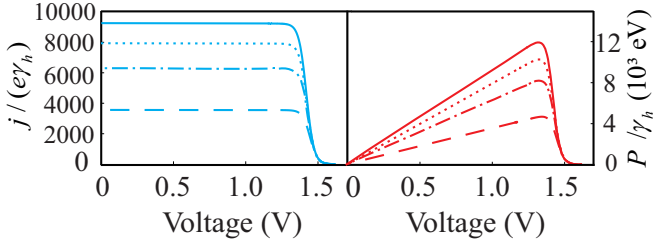}
\caption{(Color online) The current and power as a function of
effective voltage $V$ at room temperature for different numbers of
pathway. The results for one pathway are shown in dashed, two
pathways in dot-dashed, three in dotted, four in thin-solid line.
Cross-couplings corresponding to LTR2 and the second type of
cross-couplings corresponding to LTR1 are set to $0$. All the other
cross-couplings take their maximal values.}%Dorfman
\label{multiplepathways:}
\end{figure}

Electron transfer in the PSII RC has been thoroughly studied and
several charge-separation pathways were identified. In this section,
we investigate the role of multiple pathways on the behavior of
$j-V$ and $P-V$. The simulation results are presented in
Fig.~\ref{multiplepathways:}. From Fig.~\ref{multiplepathways:} we
find that as the number of pathways increases, the BQHE produces
stronger current and power. Especially, the model of two pathways,
corresponding to Fig.~\ref{energystructure1:} (b), shows a current
enhancement of $76.8\% $, comparing with the case  of only one
pathway shown in Fig.~\ref{energystructure1:} (a). Nevertheless,
with the number of pathways increase further, the enhancement rate
decreases. The current of three-pathway model shows $25.9\% $ more
than that of two-pathway model, and the current of four-pathway
model just $16.7\% $ more than that of three-pathway one.
%Fig1
The observations can be understood  as follows: the current and
power of the  BQHE  sharply depends on  the rate of transition from
the ground state $\left| b \right\rangle $ to state $\left| {{a_1}}
\right\rangle $ or $\left| {{a_2}} \right\rangle $, i.e., the decay
rate in HTR. More pathways mathematically equal to increasing the
strength of the decay rate in LTR1 and LTR2. When ${\gamma _{1h}}$,
${\gamma _{2h}}$ and ${\gamma _{12h}}$ take a fixed value, the
energy flux density is fixed. The strength of the decay rate in LTR1
and LTR2, or rather the number of pathways should match the
transport rate in HTR, otherwise, excessive pathways can not
significantly improve the transition process. We further propose
that in order to accelerate the transfer process, multiple pathways
in the BQHE (or similar structure in the natural photosynthetic
reaction center) are a better choice comparing with simply
increasing the strength of the decay rates and cross-couplings
related to LTR1 and LTR2 within one-pathway model. As shown  in
Appendix, the decay rates or cross-couplings are given by ${\gamma
_{klmn}} = \frac{{2{p^2}{V_{phn}}{g_{klp}}{g_{mnp}}}}{{\pi c}}$,
from which we know that these rates are determined by the ambient
phonon reservoir
 and the diploe moment of $\left| {{a_{1,2}}}
\right\rangle \leftrightarrow \left| b \right\rangle $ for LTR1
($\left| {{\beta _{1,2}}} \right\rangle  \leftrightarrow \left| b
\right\rangle $ for LTR2). The natural condition of reservoir is hot
and wet, and generally unstable. And the structure of RC complexes
is born fixed, resulting in constant diploe moment between different
states of pigment molecules. Therefore, these factors cannot be
utilized to enhance the reaction efficiency. Nevertheless, more
charge-separation pathways means increasing reaction channels. This
seems  easy to achieve. In the next section, considering
acceptor-to-donor charge recombination and noise-induced dephasing,
we will show that multiple pathways is still   a better choice in
contrast  with   increasing the strength of the decay rate in LTR1
and LTR2.
%Fig1
\begin{figure}[htbp]
\centering
\includegraphics[scale=0.26]{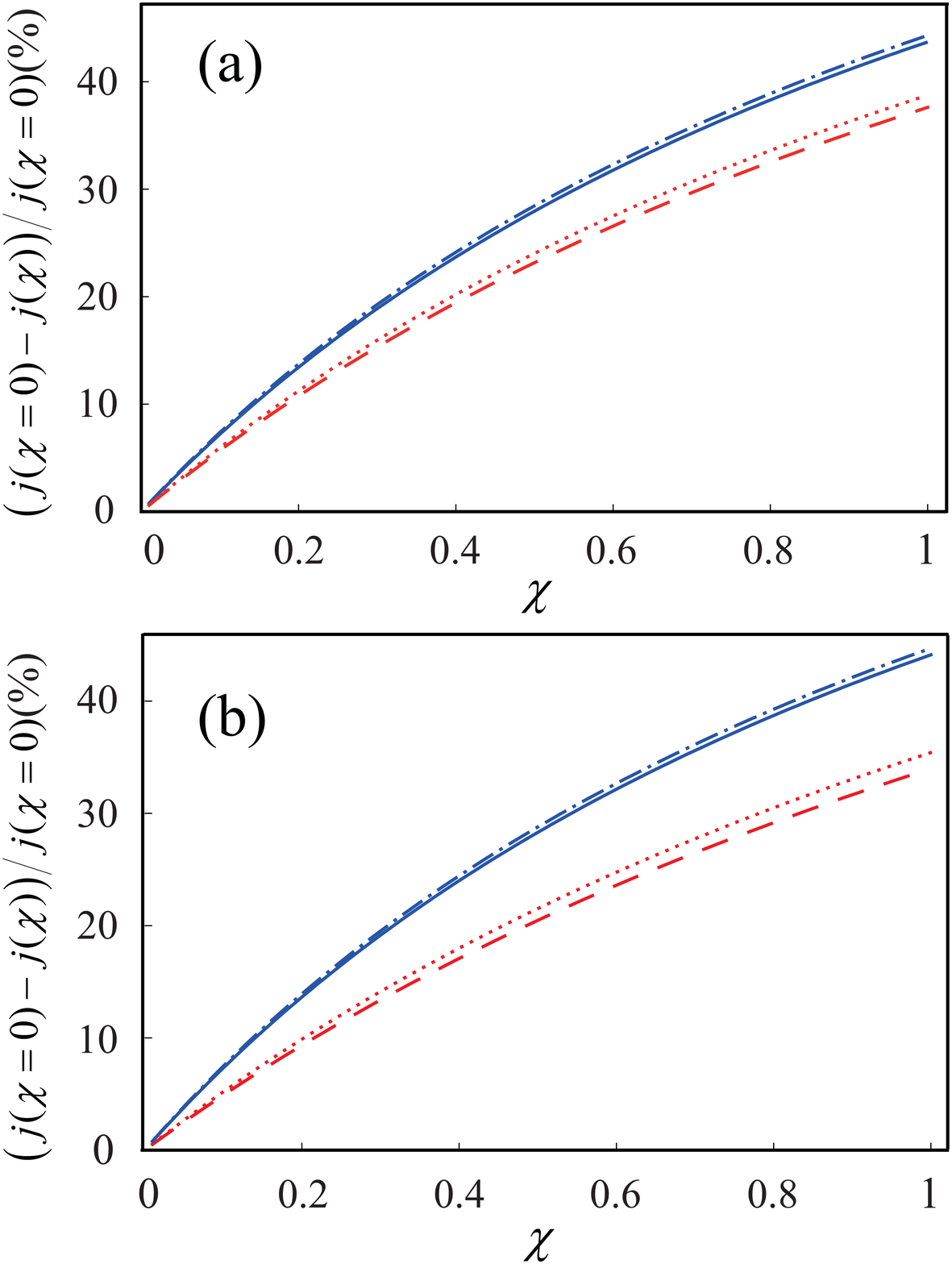}
\caption{(Color online) (a) is plotted for the current ratio defined
by  $({j(\chi  = 0) - j(\chi )})/{j(\chi  = 0)}$  as a function of
the recombination rate for one- and two-pathway models. The
blue-solid and red-dashed lines are for the maximal coherence, i.e.,
all the cross-couplings take the maximal values, except that
cross-couplings induced by LTR2 and the second type of
cross-couplings induced by LTR1 are set to $0$. The blue-dot-dashed
and red-dotted lines are for the minimal coherence, i.e., all the
cross-couplings take the minimal values 0. The blue lines stand for
one-pathway model with the strength of the decay rates in LTR1 and
LTR2 and the relaxation rate doubled and the other parameters fixed.
The red lines stand for two-pathway one. (b) is the case of one- and
three-pathway models. The blue-solid and red-dashed lines are for
the maximal coherence while the blue-dot-dashed and red-dotted lines
are for the minimal coherence. The blue lines stand for one-pathway
model with the strength of the decay rates in LTR1 and LTR2 and the
relaxation rate tripled and the other parameters fixed. The red
lines stand for three-pathway one.} \label{recombination:}
\end{figure}

\section{Effects  of recombination and dephasing}

In this section, we examine  how charge recombination at the
acceptor/donor interface and dephasing   of $\left| {{a_1}}
\right\rangle $ and $\left| {{a_2}} \right\rangle $ affects the
current and power. In Fig.~\ref{recombination:} (a), we plot the
relative change of current
as a function of    %PRL Supplement
the recombination rate $\chi $. The relative change of current  is
defined by ${{\left( {j(\chi = 0) - j(\chi )} \right)}
\mathord{\left/ {\vphantom {{\left( {j(\chi  = 0) - j(\chi )}
\right)} {j(\chi  = 0)}}} \right. \kern-\nulldelimiterspace} {j(\chi
= 0)}}$. We calculate this relative change  for both cases of
maximal coherence and minimal coherence with one and two pathways,
respectively. To determine which one, multiple pathways or large
decay rates in LTR1 and LTR2, is a better choice, we double these
decay rates and the relaxation rate in one-pathway model with the
other parameters fixed. Fig.~\ref{recombination:} (b) shows the
relative current change for both cases of maximal coherence and
minimal coherence with one and three pathways, respectively.
Similarly, we triple these decay rates and the relaxation rate in
one-pathway model with the other parameters fixed.
Fig.~\ref{recombination:} (a) and Fig.~\ref{recombination:} (b) show
that as the recombination rate increases, the current decreases
monotonically as expected. Nevertheless, the set-up with maximal
coherence can better reduce the impact of acceptor-to-donor charge
recombination. Therefore, noise-induced coherence between
$\left|{{a_1}} \right\rangle $ and $\left| {{a_2}} \right\rangle $
due to Fano interference can diminish  the influence of
recombination. Besides, we observe that the larger number of the
pathways is, the more robust the system against the recombination.
Therefore, in the presence of the additional loss mechanism due to
acceptor-to-donor charge recombination, the maximal coherence
scheme/multi-pathway set-up is more efficient than the minimal
coherence scheme/one-pathway one.

\begin{figure}[htbp]
\centering
\includegraphics[scale=0.285]{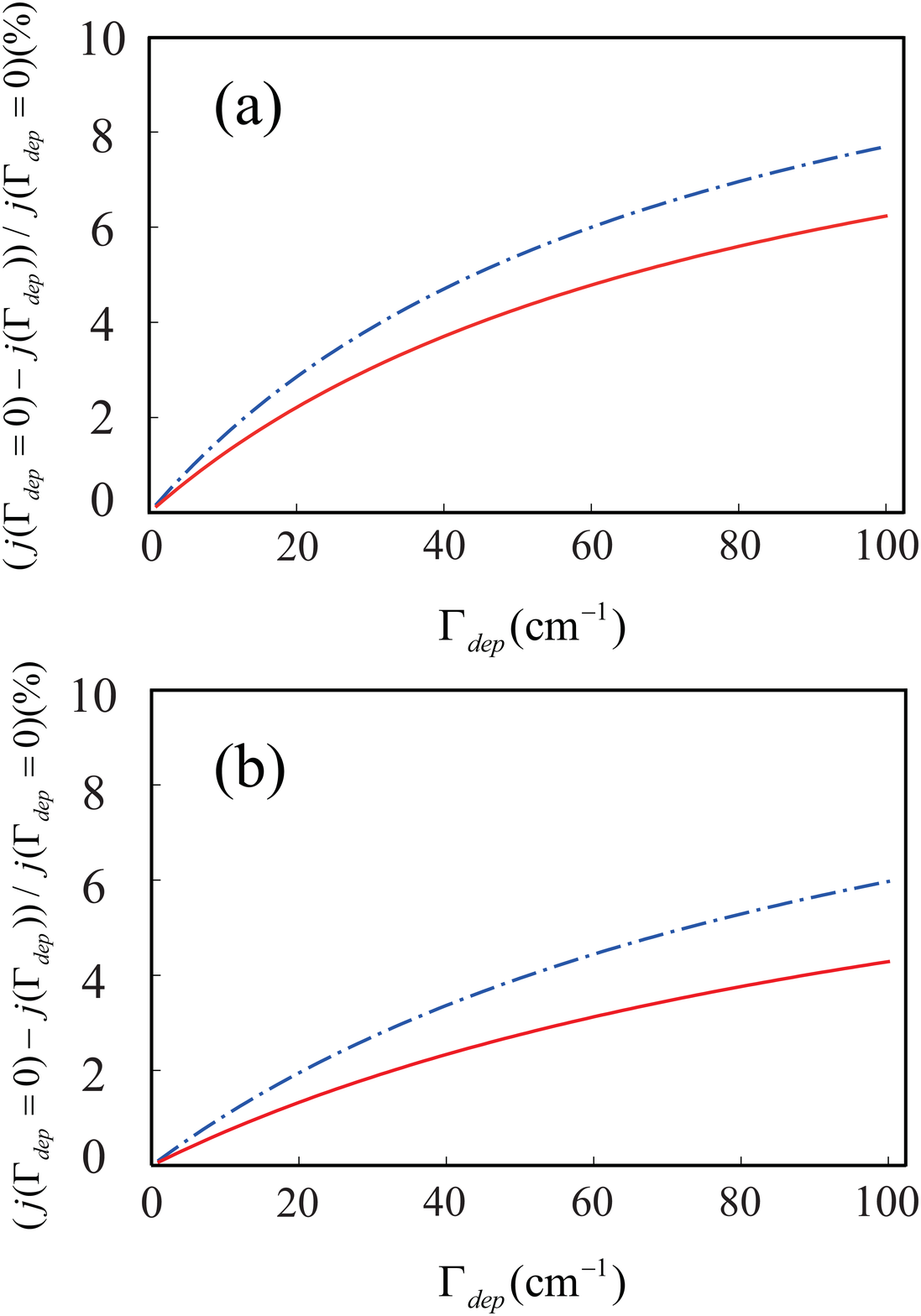}
\caption{(Color online) (a) is the relative  current change (defined
by $(j({\Gamma _{dep}} = 0) - j({\Gamma _{dep}}))/j({\Gamma _{dep}}
= 0)$) as a function of  dephasing rate. The blue-dashed line is for
one-pathway model with the strength of the decay rates in LTR1 and
LTR2 and the relaxation rate doubled and the other parameters fixed.
The red-solid line is for two-pathway one. (b) is the case of one-
and three-pathway model. The blue-dashed line is for one-pathway
model with the strength of the decay rates in LTR1 and LTR2 and the
relaxation rate tripled and the other parameters fixed. The
red-solid line is for three-pathway one. The dimensionless parameter
$\chi=0.2$ as in Table ${\rm I}$.} \label{decoherence:}
\end{figure}

Next we explore  whether, in the presence of dephasing, the
multi-pathway set-up can still benefit the current and power. We
calculate the relative change of current as a function of the
dephasing rate ${\Gamma _{dep}}$ and present the results in
Fig.~\ref{decoherence:}. The relative change of current has also
been calculated for different numbers of pathways. Similarly, to
compare the two choices of multiple pathways and increasing the
strength of the decay rate in LTR1 and LTR2, we modify these decay
rates and the relaxation rate in one-pathway model with the other
parameters fixed, just as in the case of the relaxation rate
discussed above. Observing the results, we find that the current
decreases monotonically as the dephasing rate increases. Besides,
the more pathways, the more robust against the dephasing. Therefore,
the multiple-pathway model is more efficient than the one-pathway
one even in the presence of depasing process mechanism.

%Fig1

\section{Conclusion}

In conclusion, we have studied the dynamics of two-pathway
biological quantum heat engine  and calculated the steady-state
current-voltage ($j - V$) and power-voltage ($P - V$)
characteristics. We explored the effect of various cross-couplings,
and found that only the cross-couplings describing   transitions
between high levels and the same lower energy level have effects on
the $j - V$ and $P - V$ behaviors. We explained the physics behind
this observation using the concept of  quantum beat. Furthermore we
find  that the current and power  can be increased due to Fano
interference when the upper two levels  are in a coherent
supposition.  Finally, we  calculate  the $j - V$ and $P - V$
characteristics subjected to acceptor-to-donor charge recombination
and dephasing, and show that  noise-induced quantum coherence helps
in the suppression of the  influence of the acceptor-to-donor charge
recombination.  These results  suggest that multi-pathway BQHE can
benefit  the charge separation and the light-harvesting yields, and
it is a better choice in contrast with increasing the strength of
decay rates and cross-couplings.

%%%%%%%%%%%%%%%%%%%%%%%%%%%%%%%%%%%%%%

\section{acknowledgments}
This work is supported by National Natural Science Foundation of
China (NSFC) under Grants No. 11175032, and No. 61475033.

\appendix
\section{The derivation of Eq.~(\ref{master equation})}
In this part, we will present a derivation for the master equation
used in the maintext.  We consider a  total Hamiltonian
\begin{eqnarray}
\begin{aligned}
\hat H = {{\hat H}_S} + {{\hat H}_R} + {{\hat H}_{SR}}, \label{total
Hamiltonian}
\end{aligned}
\end{eqnarray}
where the system Hamiltonian
\begin{eqnarray}
\begin{aligned}
{{\hat H}_S} = \hbar {\omega _b}\left| b \right\rangle \left\langle
b \right| + \hbar {\omega _{{a_1}}}\left| {{a_1}} \right\rangle
\left\langle {{a_1}} \right| + \hbar {\omega _{{a_2}}}\left| {{a_2}}
\right\rangle \left\langle {{a_2}} \right| , \label{system
Hamiltonian}
\end{aligned}
\end{eqnarray}
with $\hbar {\omega _{{a_i},b}} = {E_{{a_i},b}}$ $(i = 1,2)$ being
the energy of states $\left| {{a_i}} \right\rangle $, $\left| b
\right\rangle $.

${{\hat H}_R}$ represents the Hamiltonian of the high temperature
photon reservoir (HTR)
\begin{eqnarray}
\begin{aligned}
{{\hat H}_R} = \sum\limits_p {\hbar {\nu _p}\hat b_p^\dag {{\hat
b}_p}}. \label{reservoir Hamiltonian}
\end{aligned}
\end{eqnarray}
Here ${\hat b_p^\dag }$ and ${{{\hat b}_p}}$ denote the creation and
destruction operators. The interaction Hamiltonian in the
rotating-wave approximation (RWA) is
\begin{eqnarray}
\begin{aligned}
{{\hat H}_{SR}} = \hbar \sum\limits_p {{g_{1p}}{{\hat \sigma
}_{h1}}\hat b_p^\dag }  + \hbar \sum\limits_q {{g_{2q}}{{\hat \sigma
}_{h2}}{\hat b_q^\dag}}  + {\rm{H}}{\rm{.c}}{\rm{.}},
\label{Lindblad relaxation}
\end{aligned}
\end{eqnarray}
with ${{\hat \sigma }_{h1}} = \left| b \right\rangle \langle {a_1}|$
and ${{\hat \sigma }_{h2}} = \left| b \right\rangle \langle {a_2}|$
being the lower operators. The coupling constant
$g_{1p}=-\frac{{{\wp _{{a_1}b}}} \cdot \hat{\epsilonup}_{p}
\mathscr{E}_{p}}{\hbar}$ and $g_{2q}=-\frac{{{\wp _{{a_2}b}}} \cdot
\hat{\epsilonup}_{q} \mathscr{E}_{p}}{\hbar}$, where ${{\wp
_{{a_1}b}}}$ (${{\wp _{{a_2}b}}}$) is the dipole  moment of $\left|
{{a_1}} \right\rangle \leftrightarrow \left| b \right\rangle $
($\left| {{a_2}} \right\rangle \leftrightarrow \left| b
\right\rangle $ ). $\hat{\epsilonup}_{p}$ ($\hat{\epsilonup}_{q}$)
is the polarization of the field. The electric field per photon is
$\mathscr{E}_{p,q}=(\frac{\hbar
{\nu_{p,q}}}{2{\epsilonup}_{0}V})^{\frac{1}{2}}$.

It is convenient to work in the interaction picture. The
Hamiltonian, in the interaction picture, is given by
\begin{eqnarray}
\begin{aligned}
\hat V(t) = &\hbar \sum\limits_k {{g_{1p}}\hat b_p^\dag {{\hat
\sigma }_{h1}}{e^{-i({\omega _{{a_1}b}} - {\nu _p})t}}} \\
& + \hbar \sum\limits_q {{g_{2q}}\hat b_q^\dag {{\hat \sigma
}_{h2}}{e^{-i({\omega _{{a_2}b}} - {\nu _q})t}}}  +
{\rm{H}}{\rm{.c}}{\rm{.}} . \label{interaction Hamiltonian}
\end{aligned}
\end{eqnarray}
The equation of motion for the system density operator ${\hat \rho
_S}$ is
\begin{eqnarray}
\begin{aligned}
\frac{{d{{\hat \rho }_S}(t)}}{{dt}} = & - \frac{i}{\hbar
}{\rm{T}}{{\rm{r}}_R}[\hat V(t),{{\hat \rho }_S}(0) \otimes {{\hat
\rho }_R}(0)] \\
&- \frac{1}{{{\hbar ^2}}}{\rm{T}}{{\rm{r}}_R}\int_0^t {[\hat
V(t),[\hat V(t'),{{\hat \rho }_S}(t') \otimes {{\hat \rho
}_R}(0)]]dt'}  . \label{perturbation equation}
\end{aligned}
\end{eqnarray}
Substituting  $\hat V(t)$ into Eq. (\ref{perturbation equation}), we
have,
\begin{widetext}
\begin{eqnarray}
\begin{aligned}
\frac{{d{{\hat \rho }_S}(t)}}{{dt}} = & - \int_0^t {dt'} \left\{
{\sum\limits_{p,p'} {{g_{1p}}{g_{1p'}}} } \right.{e^{i({\omega
_{{a_1}b}} - {\nu _p})t - i({\omega _{{a_1}b}} - {\nu
_{p'}})t'}}{\rm{T}}{{\rm{r}}_R}\left[ {\hat \sigma _{h1}^+ {{\hat
b}_p},\left[ {\hat b_{p'}^\dag {{\hat \sigma }_{h1}},{{\hat \rho
}_S}(t') \otimes {{\hat \rho }_R}(0)} \right]} \right] \\
&+ \sum\limits_{p,q'} {{g_{1p}}{g_{2q'}}{e^{i({\omega _{{a_1}b}} -
{\nu _p})t - i({\omega _{{a_2}b}} - {\nu _{q'}})t'}}}
{\rm{T}}{{\rm{r}}_R}\left[ {\hat \sigma _{h1}^+ {{\hat b}_p},\left[
{\hat b_{q'}^\dag {{\hat \sigma }_{h2}},{{\hat \rho
}_S}(t') \otimes {{\hat \rho }_R}(0)} \right]} \right]\\
& + \sum\limits_{q,p'} {{g_{2q}}{g_{1p'}}{e^{i({\omega _{{a_2}b}} -
{\nu _q})t - i({\omega _{{a_1}b}} - {\nu _{p'}})t'}}}
{\rm{T}}{{\rm{r}}_R}\left[ {\hat \sigma _{h2}^+ {{\hat b}_q},\left[
{\hat b_{p'}^\dag {{\hat \sigma }_{h1}},{{\hat \rho
}_S}(t') \otimes {{\hat \rho }_R}(0)} \right]} \right] \\
&+ \sum\limits_{q,q'} {{g_{2q}}{g_{2q'}}{e^{i({\omega _{{a_2}b}} -
{\nu _q})t - i({\omega _{{a_2}b}} - {\nu _{q'}})t'}}}
{\rm{T}}{{\rm{r}}_R}\left[ {\hat \sigma _{h2}^+ {{\hat b}_q},\left[
{\hat b_{q'}^\dag {{\hat \sigma }_{h2}},{{\hat \rho
}_S}(t') \otimes {{\hat \rho }_R}(0)} \right]} \right]\\
& + \sum\limits_{p,p'} {{g_{1p}}{g_{1p'}}{e^{ - i({\omega _{{a_1}b}}
- {\nu _p})t + i({\omega _{{a_1}b}} - {\nu _{p'}})t'}}}
{\rm{T}}{{\rm{r}}_R}\left[ {\hat b_p^\dag {{\hat \sigma
}_{h1}},\left[ {\hat \sigma _{h1}^+ {{\hat b}_{p'}},{{\hat \rho
}_S}(t') \otimes {{\hat \rho }_R}(0)} \right]} \right] \\
&+ \sum\limits_{p,q'} {{g_{1p}}{g_{2q'}}{e^{ - i({\omega _{{a_1}b}}
- {\nu _p})t + i({\omega _{{a_2}b}} - {\nu _{q'}})t'}}}
{\rm{T}}{{\rm{r}}_R}\left[ {\hat b_p^\dag {{\hat \sigma
}_{h1}},\left[ {\hat \sigma _{h2}^+ {{\hat b}_{q'}},{{\hat \rho
}_S}(t') \otimes {{\hat \rho }_R}(0)} \right]} \right] \\
&+ \sum\limits_{q,p'} {{g_{2q}}{g_{1p'}}{e^{ - i({\omega _{{a_2}b}}
- {\nu _q})t + i({\omega _{{a_1}b}} - {\nu _{p'}})t'}}}
{\rm{T}}{{\rm{r}}_R}\left[ {\hat b_q^\dag {{\hat \sigma
}_{h2}},\left[ {\hat \sigma _{h1}^+ {{\hat b}_{p'}},{{\hat \rho
}_S}(t') \otimes {{\hat \rho }_R}(0)} \right]} \right]\\
& + \left. {\sum\limits_{q,q'} {{g_{2q}}{g_{2q'}}{e^{ - i({\omega
_{{a_2}b}} - {\nu _q})t + i({\omega _{{a_2}b}} - {\nu _{q'}})t'}}}
{\rm{T}}{{\rm{r}}_R}\left[ {\hat b_q^\dag {{\hat \sigma
}_{h2}},\left[ {\hat \sigma _{h2}^+ {{\hat b}_{q'}},{{\hat \rho
}_S}(t') \otimes {{\hat \rho }_R}(0)} \right]} \right]} \right\} .
\label{perturbation equation expansion}
\end{aligned}
\end{eqnarray}
\end{widetext}
The sum over $p$ may be replaced by an integral through
\begin{eqnarray}
\begin{aligned}
\sum\limits_p {}  \to \frac{{{V_{pht}}}}{{{\pi ^2}}}\int_0^\infty
{dp{p^2}}. \label{sum to integrate}
\end{aligned}
\end{eqnarray}
where ${{V_{pht}}}$ is the  volume. Neglecting all memory effects
and assuming that the density matrix is a slowly varying function of
time, i.e., ${\hat \rho } (t')  \approx {\hat \rho } (t)$, we obtain
the integration over time as

\begin{eqnarray}
\begin{aligned}
\int_0^\infty  {dt'{e^{i(\omega  - {\nu _k})(t - t')}}}  = \pi
\delta (\omega  - {\nu _k}). \label{time to infinity}
\end{aligned}
\end{eqnarray}

Substitute (\ref{sum to integrate}) and (\ref{time to infinity})
into (\ref{perturbation equation expansion}) and note that
$\left\langle {{{\hat b}_p}} \right\rangle  = \left\langle {\hat
b_p^\dag } \right\rangle  = 0$, $\left\langle {{{\hat b}_p}{{\hat
b}_{p'}}} \right\rangle  = \left\langle {\hat b_p^\dag \hat
b_{p'}^\dag } \right\rangle  = 0$, $\left\langle {\hat b_p^\dag
{{\hat b}_{p'}}} \right\rangle  = {{\bar n}_p}{\delta _{pp'}}$, and
$\left\langle {{{\hat b}_p}\hat b_{p'}^\dag } \right\rangle  =
({{\bar n}_p} + 1){\delta _{pp'}}$. The master equation reduces to
\begin{widetext}
\begin{eqnarray}
\begin{aligned}
\frac{{d{{\hat \rho }_S}(t)}}{{dt}} = &  \frac{{{V_{pht}}}}{{\pi
c}}\left\{ {{p^2}g_{1p}^2\left[ {\left( {{{\bar n}_p} + 1}
\right)\left( {2{{\hat \sigma }_{h1}}{{\hat \rho }_S}(t)\hat \sigma
_{h1}^\dag  - \hat \sigma _{h1}^\dag {{\hat \sigma }_{h1}}{{\hat
\rho }_S}(t) - {{\hat \rho }_S}(t)\hat \sigma _{h1}^\dag {{\hat
\sigma }_{h1}}} \right)} \right]} \right. \\
&+ {p^2}g_{1p}^2\left[ {{{\bar n}_p}\left( {2\hat \sigma _{h1}^\dag
{{\hat \rho }_S}(t){{\hat \sigma }_{h1}} - {{\hat \sigma }_{h1}}\hat
\sigma _{h1}^\dag {{\hat \rho }_S}(t) - {{\hat \rho }_S}(t){{\hat
\sigma }_{h1}}\hat \sigma _{h1}^\dag } \right)} \right] \\
&+ {p^2}g_{2p}^2\left[ {\left( {{{\bar n}_p} + 1} \right)\left(
{2{{\hat \sigma }_{h2}}{{\hat \rho }_S}(t)\hat \sigma _{h2}^\dag  -
\hat \sigma _{h2}^\dag {{\hat \sigma }_{h2}}{{\hat \rho }_S}(t) -
{{\hat \rho }_S}(t)\hat \sigma _{h2}^\dag {{\hat \sigma }_{h2}}}
\right)} \right] \\
&+ {p^2}g_{2p}^2\left[ {{{\bar n}_p}\left( {2\hat \sigma _{h2}^\dag
{{\hat \rho }_S}(t){{\hat \sigma }_{h2}} - {{\hat \sigma }_{h2}}\hat
\sigma _{h2}^\dag {{\hat \rho }_S}(t) - {{\hat \rho }_S}(t){{\hat
\sigma }_{h2}}\hat \sigma _{h2}^\dag } \right)} \right]\\
& + {p^2}{g_{1p}}{g_{2p}}\left[ {\left( {{{\bar n}_p} + 1}
\right)\left( {{{\hat \sigma }_{h1}}{{\hat \rho }_S}(t)\hat \sigma
_{h2}^\dag  + {{\hat \sigma }_{h2}}{{\hat \rho }_S}(t)\hat \sigma
_{h1}^\dag } \right)} \right]\\
& + {p^2}{g_{1p}}{g_{2p}}\left[ {{{\bar n}_p}\left( {\hat \sigma
_{h1}^\dag {{\hat \rho }_S}(t){{\hat \sigma }_{h2}} + \hat \sigma
_{h2}^\dag {{\hat \rho }_S}(t){{\hat \sigma }_{h1}}} \right)}
\right] \\
&+ {p^2}{g_{1p}}{g_{2p}}\left[ {\left( {{{\bar n}_p} + 1}
\right)\left( {{{\hat \sigma }_{h2}}{{\hat \rho }_S}(t)\hat \sigma
_{h1}^\dag  + {{\hat \sigma }_{h1}}{{\hat \rho }_S}(t)\hat \sigma
_{h2}^\dag } \right)} \right] \\
&+ {p^2}{g_{1p}}{g_{2p}}\left. {\left[ {{{\bar n}_p}\left( {\hat
\sigma _{h2}^\dag {{\hat \rho }_S}(t){{\hat \sigma }_{h1}} + \hat
\sigma _{h1}^\dag {{\hat \rho }_S}(t){{\hat \sigma }_{h2}}} \right)}
\right]} \right\}. \label{perturbation equation expansion simplify}
\end{aligned}
\end{eqnarray}
The indicator $p$ here is redefined as $p = \frac{{{\omega
_{{a_{1,2}}b}}}}{c}$. Introducing notations
\begin{eqnarray}
\begin{aligned}
{\gamma _{klh}} = \frac{{2{p^2}{V_{pht}}{g_{kp}}{g_{lp}}}}{{\pi c}},
\label{master equation h}
\end{aligned}
\end{eqnarray}
where $k,l=1,2$, $g_{k,lp}$ is the coupling constant corresponding
to transition with frequency $\omega_{a_{1,2}b}$, and the average
occupation numbers of photons ${n_h} ={\bar n}_p$, we rewrite Eq.
(\ref{perturbation equation expansion simplify}) in a simple  form
as
\begin{eqnarray}
\begin{aligned}
\frac{{d{{\hat \rho }_S}(t)}}{{dt}} = &\sum\limits_{k,l = 1,2}
{\frac{{{\gamma _{klh}}}}{2}} [({n_h} + 1)({\sigma _{hl}}{{\hat \rho
}_S}\sigma _{hk}^\dag  + {\sigma _{hk}}{{\hat \rho }_S}\sigma
_{hl}^\dag  - \sigma _{hl}^\dag {\sigma _{hk}}{{\hat \rho }_S} -
{{\hat \rho }_S}\sigma _{hk}^\dag {\sigma _{hl}}) \\
&+ {n_h}(\sigma _{hk}^\dag {{\hat \rho }_S}{\sigma _{hl}} + \sigma
_{hl}^\dag {{\hat \rho }_S}{\sigma _{hk}} - {\sigma _{hk}}\sigma
_{hl}^\dag {{\hat \rho }_S} - {{\hat \rho }_S}{\sigma _{hl}}\sigma
_{hk}^\dag )] . \label{master equation h}
\end{aligned}
\end{eqnarray}
\end{widetext}
The cross-couplings ${{\gamma _{klh}}}$ for $k \ne l$  corresponds
to the strength of the Fano interference, and ${\gamma _{kkh}} =
{\gamma _{kh}}$ denotes  the spontaneous decay rate. With this
consideration,  we finally obtain Eq. (\ref{Lindblad h}).

By the same procedure outlined for the transition $\left| b
\right\rangle \leftrightarrow \left| {{a_{1,2}}} \right\rangle $
induced by  the high temperature photon reservoir (HTR), we easily
obtain  the Lindblad term describing the transition process $\left|
b \right\rangle \leftrightarrow \left| {{\beta _{1,2}}}
\right\rangle $ induced by the second low temperature reservoir
(LTR2) in Eq. (\ref{Lindblad c2}), where ${\hat \sigma _{ck}} =
\left| b \right\rangle \left\langle {{\beta _k}} \right|$
 $(k = 1,2)$ is the  lower operator. ${n_{c2}}$ is the
average phonon numbers of LTR2. The spontaneous decay rates and the
cross-couplings are connected by,
\begin{eqnarray}
\begin{aligned}
{\Gamma _{klc}} = \frac{{2{p^2}{V_{phn}}{g_{kp}}{g_{lp}}}}{{\pi c}}.
 \label{gamma c1}
\end{aligned}
\end{eqnarray}
where the indicator $p$ is defined as
$p=\frac{\omega_{\beta_{1,2}b}}{c}$, ${g_{k,lp}}$ is the coupling
constant for transition with frequency $\omega_{\beta_{1,2}b}$
$V_{phn}$ is the phonon volume.

For the transition   $\left| {{a_{1,2}}} \right\rangle
\leftrightarrow \left| {{\alpha _{1,2}}} \right\rangle $ induced by
the first low temperature phonon reservoir (LTR1), derivation is
tedious but  similar to the presented one. The complicity comes from
the  extra charge-separated states. The Hamiltonian describing the
coupling of the system to the LTR1 in the interaction picture reads,
\begin{widetext}
\begin{eqnarray}
\begin{aligned}
\hat V(t) =& \hbar \sum\limits_p {{g_{11p}}\hat \sigma _{11}^\dag
{{\hat b}_p}{e^{i({\omega _{{a_1}{\alpha _1}}} - {\nu _p})t}}}  +
\hbar \sum\limits_q {{g_{12q}}\hat \sigma _{12}^\dag {{\hat
b}_q}{e^{i({\omega _{{a_1}{\alpha _2}}} - {\nu _q})t}}} \\
& + \hbar \sum\limits_r {{g_{21r}}\hat \sigma _{21}^\dag {{\hat
b}_r}{e^{i({\omega _{{a_2}{\alpha _1}}} - {\nu _r})t}} + \hbar
\sum\limits_s {{g_{22s}}\hat \sigma _{22}^\dag {{\hat
b}_s}{e^{i({\omega _{{a_2}{\alpha _2}}} - {\nu _s})t}}}  + }
{\rm{H}}{\rm{.c}}{\rm{.}} . \label{interaction Hamiltonian 2}
\end{aligned}
\end{eqnarray}
\end{widetext}
Following the same procedure summarized above, we obtain Eq.
(\ref{Lindblad c1}), where ${{\hat \sigma }_{kl}} = \left| {{\alpha
_k}} \right\rangle \left\langle {{a_l}} \right|$ $(k,l = 1,2)$ or
${{\hat \sigma }_{mn}} = \left| {{\alpha _m}} \right\rangle
\left\langle {{a_n}} \right|$ $(m,n = 1,2)$ is the corresponding
lower operator. ${n_{c1}}$ is the average phonon numbers of LTR1.
The spontaneous decay rates are
\begin{eqnarray}
\begin{aligned}
{\gamma _{klmn}} = \frac{{2{p^2}{V_{phn}}{g_{klp}}{g_{mnp}}}}{{\pi
c}}. \label{gamma c1}
\end{aligned}
\end{eqnarray}
with ${V_{{\rm{phn}}}}$ the phonon volume. The indicator $p$ is
defined as $p = \frac{{{\omega _{{a_i}{\alpha _i}}}}}{c}$ ($i=1,2$).
${g_{kl,mnp}}$ is the coupling constant.  ${\gamma _{klmn}}(kl =
mn)$ are the decay rates from the upper levels $\left| {{a_1}}
\right\rangle $ and $\left| {{a_2}} \right\rangle $ to the lower
levels $\left| {{\alpha _1}} \right\rangle $ and $\left| {{\alpha
_2}} \right\rangle $, respectively. The cross-couplings ${\gamma
_{klmn}}$ ($kl \ne mn$) would lead to  interferences. As the
involved levels (coupled to LTR1) here are more than those to HTR or
LTR2, there are additional  cross-couplings that do not exist in the
case of HTR or LTR2. We obtain two types of cross-couplings in the
case of LTR1. One is the same with that of HTR or LTR2, including
${\gamma _{11,21}}$ and ${\gamma _{12,22}}$. The other type includes
${\gamma _{11,12}}$, ${\gamma _{11,22}}$, ${\gamma _{12,21}}$,
${\gamma _{21,22}}$.  The first type describes transitions to the
same lower state, i.e., $\left| {{\alpha _1}} \right\rangle $ for
${\gamma _{11,21}}$ and $\left| {{\alpha _2}} \right\rangle $ for
${\gamma _{12,22}}$. But the second type  is different.

\end{document}